%% file: bcHe2_arXiv.tex
\documentclass[aps,jcp,superscriptaddress,amsmath,amssymb]{revtex4-2}

\usepackage[utf8]{inputenc}
\usepackage{amsmath}
\usepackage{comment}
\usepackage{mathtools}
\usepackage{xcolor}
\usepackage[unicode]{hyperref}
\usepackage{revsymb}
\usepackage{tikz}
\usepackage{siunitx}
\usepackage{multirow}

\hypersetup{
   unicode=true,          
   plainpages=false,
   colorlinks=true,       
   linkcolor=blue,          
   citecolor=blue,        
}
\usepackage{url}

\usepackage{dcolumn}
\newcolumntype{d}[1]{D{.}{.}{#1}}

\newcommand{\bos}[1]{\boldsymbol{#1}}
\def\Eh{E_\text{h}}
\def\cm{\text{cm}^{-1}}

\def\br{\boldsymbol{r}}
\def\iim{\text{i}}

\def\epsi{\varepsilon}

\def\nel{n_\text{e}}

\def\atSup{\text{a}\ ^3\Sigma_\text{u}^+}

\def\Eh{E_\text{h}}
\def\cm{\text{cm}^{-1}}
\def\br{\boldsymbol{r}}
\def\iim{\text{i}}
\def\Nb{N_\text{b}}
\def\ne{n_\text{el}}
\def\nnuc{N_\text{nuc}}

\def\MV{\text{MV}}
\def\Done{\text{D1}}
\def\Dtwo{\text{D2}}
\def\OO{\text{OO}}

\def\bs{\boldsymbol{s}} 

\def\hbp{\hat{\boldsymbol{p}}}
\def\hbs{\hat{\boldsymbol{s}}} 

\def\br{\boldsymbol{r}}

\def\Rnucvec{\boldsymbol{\rho}}

\def\bA{\boldsymbol{A}}
\def\hH{\hat{H}}

\def\vphi{\varphi}
\def\brho{\bos{\rho}}
\def\el{\text{el}}
\newcommand{\pd}[2]{\frac{\partial #1}{\partial #2}}

\definecolor{ao}{rgb}{0.0, 0.5, 0.0}

\usepackage{setspace}
\usepackage[unicode]{hyperref}
\usepackage{soul}
\hypersetup{
   unicode=true,          
   plainpages=false,
   colorlinks=true,       
   linkcolor=black,          
   linkcolor=blue,          
   citecolor=blue,        
   urlcolor=blue           
}
\newcommand{\hbos}[1]{\hat{\boldsymbol{#1}}}

\def\Eh{E_\text{h}}
\def\cm{\text{cm}^{-1}}

\def\br{\boldsymbol{r}}
\def\iim{\text{i}}

\def\epsi{\varepsilon}

\def\nel{n_\text{e}}

\def\ael{\text{a}}
\def\bel{\text{b}}
\def\cel{\text{c}}
\def\Bbel{\text{B}}
\def\Ccel{\text{C}}

\def\atSup{$\text{a}\ ^3\Sigma_\text{u}^+$}
\def\dtSup{$\text{d}\ ^3\Sigma_\text{u}^+$}
\def\etPg{$\text{e}\ ^3\Pi_\text{g}$} 

\def\btPg{$\text{b}\ ^3\Pi_\text{g}$}
\def\ctSgp{$\text{c}\ ^3\Sigma_\text{g}^+$}
\def\BsPg{$\text{B}\ ^1\Pi_\text{g}$}
\def\CsSgp{$\text{C}\ ^1\Sigma_\text{g}^+$}

\def\bx{\bel^{x,\Sigma}}
\def\by{\bel^{y,\Sigma}}

\def\bpm{\bel^{\pm,\Sigma}}
\def\bmp{\bel^{\mp,\Sigma}}

\def\Eh{E_\text{h}}
\def\cm{\text{cm}^{-1}}

\def\br{\boldsymbol{r}}
\def\iim{\text{i}}
\def\Nb{N_\text{b}}
\def\ne{n_\text{el}}
\def\nnuc{N_\text{nuc}}

\def\MV{\text{MV}}
\def\Done{\text{D1}}
\def\Dtwo{\text{D2}}
\def\OO{\text{OO}}

\def\bs{\boldsymbol{s}} 

\def\hbp{\hat{\boldsymbol{p}}}
\def\hbs{\hat{\boldsymbol{s}}} 

\def\Rnucvec{\boldsymbol{\rho}}
\def\brho{\boldsymbol{\rho}}

\def\el{\text{e}}

\def\bA{\boldsymbol{A}}
\def\btheta{\boldsymbol{\theta}}
\def\Pel{\mathcal{P}}

\def\hH{\hat{H}}
\def\hL{\hat{L}}
\def\vphipn{\varphi_{n'}}

\def\vphi{\varphi}
\def\nablarho{ \bos{\nabla}_{\bos{\rho}} }

\def\DMJ{\tilde{D}_{M_J\Omega}^J}
\def\DMJp{\tilde{D}_{M_J\Omega'}^J}
\def\resolvents{\hat{\mathcal{R}}_{n'} + \hat{\mathcal{R}}_{n}}
\def\mres{\hat{\mathcal{R}}}
\def\lam{\Lambda}
\def\sig{\Sigma}
\def\ome{\Omega}

\def\Nelspace{N_\mathcal{P}}

\newcommand{\bhat}[1]{\boldsymbol{\hat{#1}}}

\definecolor{ao}{rgb}{0.0, 0.5, 0.0}

\def\som{Supplementary Material}

\begin{document}

\title{%
High-Precision
Quantum Dynamics of He$_2$ over the b~$^3\Pi_\text{g}$-c~$^3\Sigma_\text{g}^+$ Electronic Subspace by including Non-adiabatic, Relativistic and QED Corrections and Couplings
}
\author{Balázs Rácsai}
\affiliation{MTA–ELTE `Momentum' Molecular Quantum electro-Dynamics Research Group,
Institute of Chemistry, Eötvös Loránd University, Pázmány Péter sétány 1/A, Budapest, H-1117, Hungary}

\author{Péter Jeszenszki}
\affiliation{MTA–ELTE `Momentum' Molecular Quantum electro-Dynamics Research Group,
Institute of Chemistry, Eötvös Loránd University, Pázmány Péter sétány 1/A, Budapest, H-1117, Hungary}

\author{Ádám Margócsy}
\affiliation{MTA–ELTE `Momentum' Molecular Quantum electro-Dynamics Research Group,
Institute of Chemistry, Eötvös Loránd University, Pázmány Péter sétány 1/A, Budapest, H-1117, Hungary}

\author{Edit Mátyus}
\email{edit.matyus@ttk.elte.hu}
\affiliation{MTA–ELTE `Momentum' Molecular Quantum electro-Dynamics Research Group,
Institute of Chemistry, Eötvös Loránd University, Pázmány Péter sétány 1/A, Budapest, H-1117, Hungary}

\date{\today}

\begin{abstract}
\noindent
Relativistic, quantum electrodynamics, as well as non-adiabatic corrections and couplings, are computed for the \btPg\ and \ctSgp\ electronic states of the helium dimer. The underlying Born-Oppenheimer potential energy curves are converged to 1~ppm ($1:10^6$) relative precision using a variational explicitly correlated Gaussian approach.
The quantum nuclear motion is computed over the \btPg--\ctSgp\ (and \BsPg--\CsSgp) 9-(12-)dimensional electronic-spin subspace coupled by non-adiabatic and relativistic (magnetic) interactions. The electron's anomalous magnetic moment is also included; its effect is expected to be visible in high-resolution experiments.
The computed rovibronic energy intervals are in excellent agreement with available high-resolution spectroscopy data, including the rovibronic \btPg-state fine structure. Fine-structure splittings are also predicted for the \ctSgp\ levels, which have not been fully resolved experimentally, yet. 
\end{abstract}

\maketitle

\noindent
This work focuses on the electronically excited states of the triplet helium dimer, which, in contrast to the very weakly bound singlet helium dimer ground state \cite{CePrKoMeJeSz12}, possesses a rich rovibronic 
level structure with intriguing magnetic properties. Triplet helium dimer states have recently attracted experimental attention for generating cold molecules and using them in precision tests of quantum electrodynamics~\cite{JaSeMe16,SeJaClAgScMe18,SeJaCaMeScMe20,VeZeKnRoBe25}. So far, the \atSup\ state of He$_2$ was used to generate He$_2^+$ rovibrational states for precision spectroscopy measurements \cite{SeJaMe16,JaSeMe16vib,JaSeMe18}. 
He$_2^+$ is a small, calculable molecular system~\cite{TuPaAd12,Ma18he2p,FeKoMa20,paper-he2p}, and thus, an ideal target for precision physics. In comparison with the H$_2$ molecule, another precision spectroscopy prototype~\cite{BeHoHuChSaEiUbJuMe19,PuKoCzPa19,CoDiUb23,PaKo24}, the magnetic properties of the helium dimer (and its cation) can potentially offer a combination of versatile experimental tools and techniques for cooling, quantum state preparation, and measurement.
Furthermore, the electronically excited helium dimer states have recently been linked to plasma kinetics~\cite{XuLuZhGuSh24}.
In addition to ongoing experimental work~\cite{JaSeMe16,SeJaClAgScMe18,SeJaCaMeScMe20,VeZeKnRoBe25}, old experimental data on the electronic-vibrational-rotational level structure and even fine-structure splittings of some transitions are available, \emph{e.g.,} \cite{Gi65,Gi365,GiBa70,GiGi80,
FoBeCo98,RoBrBeBr88,LoKeBj89,BrGi71,BrHa79,KrBj90,HaNoBj95}, which have remained largely unexplained by \emph{ab initio} molecular quantum theory to date.
It may be orienting to mention that the \atSup\ state of the He$_2$ Rydberg molecule is the $2s$ 
state, and the \btPg\ and \ctSgp\ are the closest-lying gerade states, corresponding to $p$ character.

Early theoretical and computational work includes studies by valence bond theory computations \cite{BuDa52,Mu64,GuGo75}. Then, the multiconfigurational self-consistent field (MCSCF) approach, with special, augmented basis sets, was used \cite{WaStJa81,SuLiSiSiJoSh83,ChJeYaLe89,Ya89,Mi03}, though the results lag behind even the old experimental data. 
More recent computational work employed the Multireference Configuration Interaction (MRCI) method \cite{BjMiPaRo98}, the Equation-of-Motion Coupled Cluster method~\cite{NiKrPrViWi2019}, MRCI combined with neural networks~\cite{XuLuZhGuSh24}, and the R-matrix method~\cite{EpMoChTe24}. Furthermore, there is a single variational explicitly correlated Gaussian (ECG) computation~\cite{PaCaBuAd08}, which has been the most accurate so far. Despite all these efforts, a considerable gap still exists between theory and experiment for this simple Rydberg molecule, which is the smallest excimer. 

This work, part of a series of variational explicitly correlated Gaussian computations carried out by our group~\cite{paper-he2p,paper1,paper3} aims to provide computational results sufficiently accurate to guide, help interpret, perhaps even challenge, modern experiments on this molecular physics prototype.

%
%
For direct comparison with experimental spectroscopy transitions and energy intervals, we start out from the non-relativistic molecular Schrödinger equation
\begin{align}
  \hat{H} \Psi(\br,\brho) &= E \Psi(\br,\brho) 
  \label{eq:molSch-eq1}
\end{align}
with the translationally invariant Hamiltonian
\begin{align}
    \hat{H}
    =
    -
    \frac{1}{2 \mu}
    \bos{\Delta}_{\brho}
    +
    \hat{H}_{\el}
    -
    \frac{1}{8 \mu}
    \hat{\bos{P}}_{\el}^2
    \; ,
    \label{eq:molHam}
\end{align}
where $\mu=M_\text{nuc}/2$ is the reduced mass, $\brho=\bos{R}_1-\bos{R}_2$ is the internuclear position vector, and
$\hat{\bos{P}}_{\el}=-\iim\sum_{i=1}^{\nel} \bos{\nabla}_{\br_i}$ is the total (linear) momentum of the $\nel$ electrons. 
$\hat{H}_{\el}$ is the electronic Hamiltonian, for which we (numerically) solve the
\begin{align}
  \hat{H}_{\el} \varphi_n = U_n \varphi_n
  \label{eq:elSch}
\end{align}
Schrödinger equation (with $\langle \varphi_{n'}|\varphi_n \rangle = \delta_{n'n}$) to obtain the $\varphi_n$ electronic states relevant for the experiments. 
We perform the electronic structure computations in the body-fixed (BF) frame in which the $\brho$ internuclear vector defines the $z$ axis. The electronic state is characterised by its total spin, 
$\hat{S}^2 \vphi_n=S(S+1) \vphi_n$, and the $z$ BF axis being the quantisation axis for the orbital and spin angular momenta of the electrons, $\hat{L}_z \vphi_n = \lam \vphi_n$ and 
$\hat{S}_z \vphi_n = \sig \vphi_n$.

In order to numerically solve the Schrödinger equation of the molecular Hamiltonian, Eqs.~\eqref{eq:molSch-eq1}--\eqref{eq:molHam}, we write the molecular wave function as a linear combination of products of electronic, rotational, and vibrational functions,
\begin{align}
  \Psi(\br,\Rnucvec)
  =
  \sum_{n \in \Nelspace}
  \sum_{\ome =-J}^J
  \sum_{k} 
    c_{n,k}^{\Omega}
    \varphi_n(\br, \Rnucvec) 
    \DMJ (\theta, \phi, \chi)
    \frac{1}{\rho}
    g_{k}(\rho) \; ,
  \label{eq:mol-wave-fun}
\end{align}
where $n=(l,S,\lam,\sig)$ collects the relevant electronic state labels and quantum numbers 
and it is summed over all electronic states included in the $\Pel$ `active' subspace (for which the indices are collected in the set $\Nelspace$). $g_k(\rho)$ are the vibrational basis functions.
The (normalized) $\DMJ(\theta, \phi, \chi)$ Wigner D matrices describe the rotation of the body-fixed frame $(x,y,z)$ (rotating diatom with BF spin and orbital angular momenta) with respect to the laboratory-fixed system $(X,Y,Z)$, where the $\ome = \lam + \sig$ and $M_J$ quantum numbers denote the projection of the total rotational plus electronic angular momentum, onto the $z$ and $Z$ axes, respectively,
$\hat{J}_z \DMJ = \ome \DMJ$, $\hat{J}_Z \DMJ = M_J \DMJ$, and $\hat{J}^2 \DMJ = J(J+1) \DMJ$.
The $\hbos{J}$ total angular momentum operator acts on the $(\theta, \phi, \chi)$ Euler angles of the Wigner D matrices, while the $\hbos{S}$ and $\hbos{L}$ operators act on the electronic spin and spatial degrees of freedom. 
The helium-4 nuclei have zero spin, so the nuclear spin angular momentum is considered only for the spin statistics (of bosonic nuclei).

We insert the Ansatz of Eq.~\eqref{eq:mol-wave-fun} into Eq.~\eqref{eq:molSch-eq1}, multiply both sides from the left by the product of an electronic and a Wigner D element, and integrate for the electronic coordinates and the Euler angles to arrive at the following coupled radial (vibrational) equation,
\begin{align}
    E
    &\sum_{k} 
    c_{n',k}^{\Omega'}
    \frac{1}{\rho}
    g_{k}
    =
    \sum_{n \in \Nelspace}
    \sum_{\ome = -J}^J
    \sum_{k} 
    c_{n,k}^{\Omega} 
    \langle
    \vphi_{n'} \DMJp |
    -
    \frac{1}{2 \mu}
    \bos{\Delta}_{\brho}
    +
    \hat{H}_{\el}
    -
    \frac{1}{8 \mu}
    \hat{P}_{\el}^2
    | \varphi_n
    \DMJ \rangle
    \frac{1}{\rho}
    g_{k}
    \; .
    \label{eq:cpl-rad-eq-1}
\end{align}
For practical computations, we rewrite the nuclear Laplacian using the $\hbos{R}$ rotational angular momentum operator as
\begin{align}
    \bos{\Delta}_{\brho}
    &=
    \frac{1}{\rho^2}
    \frac{\partial}{\partial \rho}
    \left(
    \rho^2
    \frac{\partial}{\partial \rho}
    \right)
    -
    \frac{1}{\rho^2}
    \hat{\bos{R}}^2
    \nonumber \\
    &=
    \frac{1}{\rho^2}
    \frac{\partial}{\partial \rho}
    \left(
    \rho^2
    \frac{\partial}{\partial \rho}
    \right)
    \nonumber \\
    &-
    \frac{1}{\rho^2}
    \Big[
    (\hat{J}^2 - \hat{J}_z^2)
    +
    (\hat{L}^2_x + \hat{L}^2_y)
    +
    (\hat{S}^2 - \hat{S}_z^2)
    \nonumber \\
    &-
    (\hat{J}^+\hat{S}^- + \hat{J}^-\hat{S}^+)
    -
    (\hat{J}^+\hat{L}^- + \hat{J}^-\hat{L}^+)
    \nonumber \\
    &+
    (\hat{S}^+\hat{L}^- + \hat{S}^-\hat{L}^+)
    \Big]
    \label{eq:nuc-Lap-JLS}
    \; .
\end{align}
The $\hat{O}^{\pm} = \hat{O}_x\pm \iim \hat{O}_y$ body-fixed ladder operator matrix elements are calculated according to established relations~\cite{LBFi04book}
$
    \langle 
      D^J_{M_J \Omega'} |
      \hat{J}^\pm 
      D^J_{M_J \Omega} 
    \rangle
    =
    \delta_{\Omega',\Omega \mp 1}
    C_{J\ome}^\mp
$, 
$
    \langle 
      \vphipn | 
        \hat{S}^\pm
      \vphi_n
    \rangle
    =
    \delta_{\Sigma',\Sigma \pm 1}
    C_{S\sig}^\pm
$
with $C_{AB}^{\pm} = [A(A+1)-B(B\pm 1)]^{1/2}$. The orbital angular momentum integrals for the electronic states of molecules are computed numerically, 
$
    \langle 
      \vphipn | 
        \hat{L}^\pm
      \vphi_n
    \rangle
    =
    \delta_{\lam',\lam\pm1}
    \langle 
      \vphi_{n'} | 
        [\hat{L}_x \pm \iim \hat{L}_y]
      \vphi_{n}
    \rangle
    \; .
    \label{eq:angmomL}
$
For computational feasibility, we explicitly include only the qualitatively important electronic states in the $\Pel$ subspace. However, for rovibronic computations of spectroscopic accuracy, it is necessary to account for the remaining (small) effect of the $(1-\Pel)$ electronic Hilbert space. By perturbation theory (contact transformation), this leads to the following correction~\cite{MaTe19,MaFe22nad} to the right side of Eq.~\eqref{eq:cpl-rad-eq-1}, 
\begin{align}
    \frac{1}{2 \mu^2}
    \sum_{n} 
    \sum_{\ome}
    \langle \DMJp |
    \bos{\nabla}_{\bar{\bos{\rho}}}
    \cdot
    \langle
    \bos{\nabla}_{\bar{\bos{\rho}}}
    \varphi_{n'} |
    \mres
    | \nablarho \varphi_n \rangle
    \nablarho
    \cdot
    | \DMJ \rangle
    \; ,
    \label{eq:nmass-corr-1}
\end{align}
where $\bar{\bos{\rho}}=\brho=\bos{R}_1-\bos{R}_2$ was introduced only to help label the corresponding factors of the two scalar products unambiguously.
The correction contains the reduced resolvent of the electronic Hamiltonian,
\begin{align}
    2\mres
    =
    \resolvents
    =
    \left[ 
    (\hat{H}_{\el} - E_{n'})^{-1}
    +
    (\hat{H}_{\el} - E_n)^{-1}
    \right]
    P^\perp
    \; ,
\end{align}
where $P^\perp=1 - \sum_{n \in \Pel} | \vphi_n\rangle\langle \vphi_n|$ projects out the electronic states explicitly coupled in Eq.~\eqref{eq:mol-wave-fun}.
For the \ctSgp\ electronic state with $\lam=0$, Eq.~\eqref{eq:nmass-corr-1} leads to the following contributions to the reduced rotational (x=r) and vibrational (x=v) masses
\begin{align}
  \frac{1}{2\mu^\text{x}_n(\rho)}
  =
  \frac{1}{2\mu}
  \left[%
    1 - \frac{\delta m^\text{x}_n(\rho)}{2\mu}
  \right]
  \approx
  \frac{1}{2\mu + \delta m^\text{x}_n(\rho)}
\end{align}
with
\begin{align}
  \delta m_n^\text{x}(\rho)
  =
  \left\langle 
    \pd{\vphi_n}{\rho_{a_\text{x}}} |
    (\hat{H}_\el-E_n)^{-1} P^\perp
    | \pd{\vphi_n}{\rho_{a_\text{x}}}
  \right\rangle \; ,
  \label{eq:vibrot-nadmass}
\end{align}
$a_\text{x}=z$ for $\text{x}=\text{v}$  and $a_\text{x}=x/y$ for $\text{x}=\text{r}$.
We note that the vibrational non-adiabatic mass (perturbatively) accounts for all homogeneous ($\Delta \Lambda=0$) non-adiabatic couplings of the state with all other states of the same spatial symmetry (not explicitly, variationally coupled in the rovibronic treatment). The rotational mass accounts for the heterogeneous coupling ($\Delta \Lambda=\pm 1$) for all states not explicitly coupled within the variational treatment. It is important to note that only the effect of distant (non-crossing, separated by a finite gap \cite{MaTe19}) states can be accounted for in this manner; crossing or close-lying electronic states must be coupled variationally (as the c and b states in the present system). Numerical illustrations for electronically excited hydrogenic states and vibronic states over coupled subspaces have been reported in Refs.~\citenum{FeMa19HH} and \citenum{MaFe22nad}.
The b-state vibrational mass correction was also computed using Eq.~\eqref{eq:vibrot-nadmass}.
At the same time, the b-state rotational mass and further contributions from the b-c coupling through the kinetic energy correction term, Eq.~\eqref{eq:nmass-corr-1}, are neglected here and left for future work.

In the rovibronic Schrödinger equation, Eq.~\eqref{eq:cpl-rad-eq-1}, we correct the electronic energy (PEC) with spin-independent (sn) and spin-dependent (sd) relativistic and QED corrections and couplings by adding the terms ($\alpha$ labelling here the fine-structure constant), 
\begin{align}
      \alpha^2
      \langle %
         \vphipn | %
         \hat{H}_{\text{sn}}^{(2)} + \hat{H}_{\text{sd}}^{(2)} 
         | \vphi_n
      \rangle 
      +
      \alpha^3 [%
      E^{(3)}_{\text{sn},\vphipn,\vphi_n} 
      +
      \langle %
         \vphipn | 
         \hat{H}^{(3)}_{\text{sd}}
         | \vphi_n
      \rangle          
      ] \; .
\end{align}
The spin-independent corrections are, in short, the centroid relativistic (QED) energy corrections, whereas the spin-dependent corrections can be understood as the terms responsible for magnetic (dipolar and contact) interactions.  The relativistic and QED corrections are computed in the non-relativistic quantum electrodynamics (nrQED) framework; the detailed expressions are reiterated in the \som.
For further details regarding the electron-spin ($\Sigma$) dependent terms, please see also Refs.~\citenum{paper1,paper3}; the centroid relativistic corrections are evaluated according to Refs.~\citenum{RaFeMaMa24}. 

All in all, we arrive at the final form of the coupled radial equation,
\begin{align}
   &E
    \sum_{k} 
    c_{n',k}^{\Omega'}
    g_{k}
  =
    \sum_{n \in \Nelspace}
    \sum_{k} 
    c_{n,k}^{\Omega}
    \Big\{
    \delta_{\lam'\lam}
    \delta_{\sig'\sig}
  \nonumber \\
  &\Big[
    -
    \frac{\text{d}}{\text{d} \rho}
    \frac{1}{2 \mu_n^{\text{v}}}
    \frac{\text{d}}{\text{d} \rho}
    +
    \frac{J(J+1)-\ome^2}{2 \mu_n^{\text{r}}\rho^2}
    +
    \frac{S(S+1)-\sig^2}{2 \mu_n^{\text{r}}\rho^2}
    +
    W_n
    \Big]
    \nonumber
    \\
    &-
    \frac{1}{2\mu_n^{\text{r}} \rho^2}
    \Big[ 
    \delta_{\lam'\lam}
    \delta_{\sig'\sig\pm1}
    C_{J\ome}^\pm C_{S\sig}^\pm
    +
    \delta_{\lam'\lam\pm1}
    \delta_{\sig'\sig}
    C_{J\ome}^\pm
    \langle \vphi_{n'} |
    \hat{L}^\pm \vphi_n
    \rangle
    \Big]
    \nonumber
    \\
    &+
    \frac{1}{2\mu^{\text{r}}_n \rho^2}
    \delta_{\lam'\lam\pm1}
    \delta_{\sig'\sig\pm1}
    C_{S\sig}^\pm
    \langle \vphi_{n'} |
    \hat{L}^\pm \vphi_n
    \rangle
    +
      \alpha^2
      \langle %
         \vphipn | %
         \hat{H}^{(2)}
         | \vphi_n
      \rangle 
    \nonumber
    \\
    &+
      \alpha^3 [E^{(3)}_{\text{sn},\vphipn,\vphi_n}+E^{(3)}_{\text{sd},\vphipn,\vphi_n}]
    \Big\}
    g_{k}  
    \; ,
    \label{eq:cplradeqJ}
\end{align}
where the BO potential energy curve with the diagonal Born-Oppenheimer (DBOC) or adiabatic correction is 
\begin{align}
    W_n
    &=
    U_{n}
    -
    \frac{1}{8\mu} 
    \langle 
    \vphi_n | 
    \hat{P}^2_{\el}
    \vphi_n
    \rangle
    -
    \frac{1}{2\mu}
    \langle 
    \vphi_n | 
    \frac{\partial^2}{\partial \rho^2}
    \vphi_n
    \rangle
\nonumber \\
    &+
    \frac{1}{2\mu\rho^2}
    \langle 
    \vphi_n | 
    \hat{L}^2_x + \hat{L}^2_y
    | \vphi_n
    \rangle
    \; .
\end{align}
We neglected the effect of the non-adiabatic contact transformation on the relativistic and QED terms, and \emph{vice versa,} the Foldy-Wouthuysen transformation on the nuclear motion operators. The corresponding terms would account for the combined relativistic, QED and non-adiabatic coupling, and their consideration is left for future work. 

Furthermore, we note that for the smallness of the magnetic splittings, we also performed computations without considering couplings due to the electron spin. In this case, it is convenient to use the total `spatial' angular momentum, $\hat{\bos{N}}=\hat{\bos{R}}+\hat{\bos{L}}$, and then, we use the Wigner $\tilde{D}_{M_N\lam}^{N}$ matrices to describe the spatial rotation of the diatom with orbital angular momentum in the BF frame, 
$\hat{N}_z \tilde{D}_{M_N\lam}^{N}=\lam \tilde{D}_{M_N\lam}^{N}$,
$\hat{N}_Z \tilde{D}_{M_N\lam}^{N}=M_N \tilde{D}_{M_N\lam}^{N}$, and 
$\hat{N}^2 \tilde{D}_{M_N\lam}^{N}=N(N+1) \tilde{D}_{M_N\lam}^{N}$.
Then, the coupled radial equation simplifies to 
\begin{align}
    E
    \sum_{k} 
    &c_{n',k}^{\Lambda'}
    g_{k}
    =
    \sum_{n \in\Nelspace}
    \sum_{k} 
      c_{n,k}^{\Lambda}
      \Big\{
        \delta_{\lam'\lam}
    \nonumber \\
    &\Big[%
    -
    \frac{\partial}{\partial \rho}
    \frac{1}{2 \mu_n^{\text{v}}}
    \frac{\partial}{\partial \rho}
    +
    \frac{N(N+1)-\lam^2}{2 \mu_n^{\text{r}}\rho^2}
    +
    W_n
    \Big] %
    \nonumber \\
    &-
    \frac{1}{2\mu_n^{\text{r}} \rho^2}
    \delta_{\lam'\lam\pm1}
    C_{N\Lambda}^\pm
    \langle \vphi_{n'} |
    \hat{L}^\pm \vphi_n
    \rangle
    \nonumber
    \\
    &+
      \alpha^2
      \langle %
         \vphipn | %
         \hat{H}^{(2)}
         | \vphi_n
      \rangle 
    +
      \alpha^3 E^{(3)}_{\text{sn},\vphipn,\vphi_n}
    \Big\}
    g_{k}  
    \; .
    \label{eq:cplradeqN}
\end{align}
For rovibrational computations on a single electronic state,
we solve the radial equation of Refs.~\cite{paper-he2p} and \cite{paper1} without and with considering fine-structure (zero-field splitting) effects, respectively.

The outlined theoretical framework (with further details in the \som) benefited from reading several pioneering spectroscopy and ab initio studies. Most importantly, we mention the DUO program documentation~\cite{YuLoTeSt16} about a modern formalism using ab initio computations for astrophysical applications of diatomic molecules including non-adiabatic and spin-dependent couplings. We have also studied the series of work on the non-adiabatic couplings in the electronically excited states of the hydrogen molecule by Dressler, Wolniewicz, and co-workers \cite{QuDrWo90,YuDr94} and even the early work of Ko{\l}os and Wolniewicz \cite{KoWo63}. The book by Lefebvre-Brion and Field~\cite{LBFi04book} and the book by Brown and Carrington~\cite{BrCa03} served as rich sources of ideas and information.

%
%
\begin{figure*}
  \includegraphics[scale=1.]{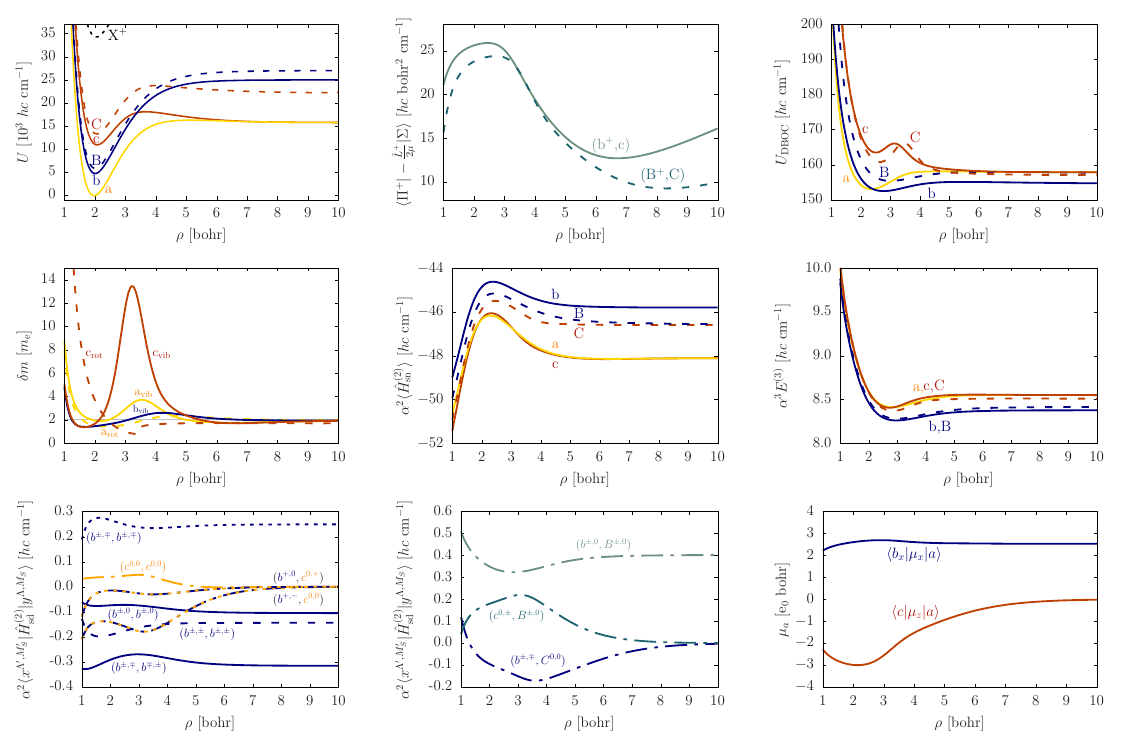}
  \caption{%
    He$_2$ b, c, B, and C electronic states: BO PEC, non-adiabatic coupling, DBOC correction, vibrational and rotational non-adiabatic mass correction, leading-order relativistic and QED correction curves computed in this work. The a-state results are reproduced from   Ref.~\citenum{paper1}, and the bottom of the X$^+\ ^2\Sigma_\text{u}^+$ BO PEC of He$_2^+$ \cite{paper-he2p} is plotted for comparison.
    The spin-dependent relativistic corrections and couplings ($\hat{H}_\text{sd}$), for the b, c, B, and C electronic states are also shown, computed in Ref.~\cite{paper3}. All non-vanishing non-adiabatic and relativistic QED couplings, computed as part of Ref.~\citenum{paper3}, are plotted (see also Tables~S6-S7).
    The electric dipole ($\mu_a$) transition moments connecting the b and c states with the a state, computed in this work, are also shown. 
    \label{fig:allcurves}
  }
\end{figure*}

The $J$ and $N$ radial equations, Eqs.~\eqref{eq:cplradeqJ} and \eqref{eq:cplradeqN}, are used with the PECs, correction and coupling curves evaluated numerically at a series of nuclear geometries (and then interpolated through the points). These quantities are shown in Figure~\ref{fig:allcurves}; they were computed numerically with the in-house developed computer program, QUANTEN, and are briefly explained in the following paragraphs.
First, we solved the electronic Schrödinger equation, Eq.~\eqref{eq:elSch}, in a non-linear variational procedure. The electronic wave function is written as a linear combination
of anti-symmetrized products of $f(\br;\bos{A}_i,\bos{s}_i)$ floating explicitly correlated Gaussian functions (fECG) and $\chi$ spin functions,
\begin{align}
  &\vphi_n(\br) 
  =
  \sum_{i=1}^{N_\text{b}}
    d_i
    \hat{\mathcal{P}}_\text{G} \hat{\mathcal{A}}_\text{S}
    \left\lbrace
      f(\br;\bA_i,\bs_i)
      \chi_{(S,\Sigma)}(\bos{\theta}_i) 
    \right\rbrace \; ,
  \\
  &f(\br;\bA_i,\bs_i)
  =
  \exp%
    \left[%
      (\br-\bs_i)^\text{T} \underline{\bos{A}}_i (\br-\bs_i)
    \right] \; ,   
\end{align}
where the $\underline{\bos{A}}_i=\bos{A}_i\otimes\bos{I}_3$ matrix is a real, symmetric, positive definite matrix with the $\bos{I}_3\in\mathbb{R}^{3\times3}$ unit matrix and the $\bos{A}_i\in\mathbb{R}^{\nel\times\nel}$ matrix; $\bs_i\in\mathbb{R}^{3\nel}$ is a shift vector corresponding to the centre of the fECG function. $\chi_{(S,\Sigma)}(\bos{\theta}_i)$ is an electronic spin function constructed according to Refs.~\cite{paper1,paper3}.
The $\bA_i,\bs_i$ and $\btheta_i$ parameters were generated according to the stochastic variational method \cite{SuVaBook98,MaRe12} and refined with the Powell approach~\cite{Po04}. 
The $\hat{\mathcal{A}}_\text{S}$ electronic anti-symmetrization operator ensures fulfilment of the Pauli principle for the electronic part, while $\hat{\mathcal{P}}_\text{G}$ projects onto the selected irreducible representation (irrep) of the $D_{\infty\text{h}}$ point group. 

In this work, we computed the \ctSgp\ and the \btPg\ ($x$ component in the Cartesian representation) electronic states of the helium dimer. First, fECG basis functions were generated and optimised (Table~S1) at $\rho=2$~bohr. 
Convergence tests and comparisons with standard electronic structure computations are collected in the~\som. Based on these tests, the basis sets used for the b and c electronic energies were converged to 3--5~$\cm$ (as an upper bound to the exact value) at $\rho=2$~bohr.

The BO PECs for the b and c electronic states were generated by small, consecutive displacements of $\rho$ followed by rescaling the $\bos{s}_i$ vectors and full reoptimization (in repeated Powell cycles) of all non-linear basis parameters~\cite{CeRy95,FeMa19HH,FeKoMa20,FeMa22h3}. This PEC generation procedure was started from $\rho=2$~bohr with a $\pm$0.1~bohr step size over the [1,10]~bohr interval. This consecutive rescaling-refining procedure allowed us to approximately conserve the absolute error along a series of neighbouring nuclear configurations, and thus, the relative properties (here: rovibrational excitation energies) are expected to be (1-2 orders of magnitude) more precise than the absolute electronic energies. 
Along the PECs, the electronic wave functions obtained from the variational fECG procedure were used to compute corrections and couplings. 

Regarding the DBOC and non-adiabatic mass corrections, the
$\langle \pd{\vphipn}{\rho} | \pd{\vphipn}{\rho} \rangle$ and $\langle \vphipn | \pd{\vphi_n}{\rho} \rangle$ integrals were computed by finite differences, in which the optimised $\bos{s}_i$ fECG centres were rescaled in proportion to the (tiny) $\rho\pm \delta \rho$ variation.
The electronic orbital angular momentum operators
$\langle \vphipn | \hat{L}_a \vphi_n \rangle$ and $\langle \vphipn | \hat{L}^2_a \vphi_n \rangle\ (a=x,y)$ were computed using the analytic integrals in QUANTEN.
Auxiliary basis sets for the non-adiabatic vibrational and rotational mass corrections, Eq.~\eqref{eq:vibrot-nadmass}, were optimised by minimisation of the target quantities using the procedure of Ref.~\cite{MaFe22nad}.

Regarding the relativistic (spin-independent, centroid) corrections, we employed the numerical Drachmanization approach described in Ref.~\cite{RaFeMaMa24} (Table~S3 presents convergence tests).
The QED (centroid) corrections were calculated using the regularised Dirac delta expectation values already computed for the relativistic corrections, the Araki--Sucher (AS) term was computed with the integral transformation technique~\cite{PaCeKo05,JeIrFeMa22} for the c state (and this AS curve was used also for the b, B, and C states; see \som). For the Bethe logarithm, we use the ion-core approximation and the values of the ground electronic state of He$_2^{3+}$ \cite{FeKoMa20}. 

The computation of spin-dependent relativistic couplings with fECGs was recently implemented in QUANTEN. The methodology is reported in a separate paper~\cite{paper3}, including extensive convergence tests for the b- and c-state couplings. The leading-order QED corrections to the spin-dependent couplings are due to the electron's anomalous magnetic moment, which is also accounted for; the technical and computational details are in Refs.~\citenum{paper3,paper1}. We had to pay attention to the phase of the electronic wave function; it was fixed for the computation of all (non-adiabatic, relativistic and QED) couplings, and the phase of the (real-valued) electronic wave functions at neighbouring PEC points was chosen so that their overlap is closer to $+1$, than to $-1$.
For future computations (and to aid the experimental work), the electric dipole transition moments to the \atSup\ state were also computed as a function of the $\rho$ internuclear distance and are shown in Fig.~\ref{fig:allcurves}.
The total computing time used to prepare this paper (including electronic structure optimisation and all corrections and couplings, partly reported in Ref.~\citenum{paper3}) is estimated to be 150~kCPU~hours.

QUANTEN uses the Cartesian representation of the basis and wave functions labelled as $\varphi_n^{a,\Sigma} \ (a=x,y,0(z)$), in short:
\begin{align}
  \mathcal{B}^{(xyz)}: 
  (b^{x,-1}, b^{x,0}, b^{x,1},
   b^{y,-1}, b^{y,0}, b^{y,1},
   c^{0,-1}, c^{0,0}, c^{0,1}, 
   B^{x,0},B^{y,0},C^{0,0}) \; .
\end{align}
For the rovibrational computations, we transform all quantities from the Cartesian to the spherical representation, in which the electronic wave functions are labelled as $\varphi_n^{\Lambda,\Sigma}$, in short:
\begin{align}
  \mathcal{B}^{(-1,0,1)}: 
  (b^{-1,-1}, b^{-1,0}, b^{-1,1},
   b^{+1,-1}, b^{1,0}, b^{1,1},
   c^{0,-1}, c^{0,0}, c^{0,1}, 
   B^{+,0},  B^{-,0},C^{0,0}) 
\end{align}
with the spherical representation for the $\Pi$ states ($b$ and $B$) constructed as:
$\Pi^{\pm,M_S} = \frac{1}{\sqrt{2}}\left[\Pi^{x,M_S} + \iim \Pi^{y,M_S} \right]$
Further calculations are collected in the \som.
The non-vanishing relativistic (QED) spin-dependent matrix elements (all $\Delta\Omega=0$)
and non-adiabatic ($\Delta\Lambda=\pm 1$; $\Delta\Lambda=0$ not relevant for the present system) coupling matrix elements used in the coupled equations are collected in Tables~S7 and S8.

The coupled radial equations, Eqs.~\eqref{eq:cplradeqJ} or \eqref{eq:cplradeqN}, are solved using the discrete variable representation (DVR) constructed with $L^{(\alpha)}_n$ Laguerre polynomials ($\alpha=2$), similarly to Refs.~\citenum {Ma18he2p,FeKoMa20} (see also \cite{AvMa19}). 
At the end of the rovibrational and rovibronic computations, we retain the states, for comparison with experiment, that are totally symmetric under the exchange of the two $^4$He$^{2+}$ nuclei (spin-0 bosons) \cite{BrCa03}.

\begin{figure*}
  \includegraphics[scale=1.]{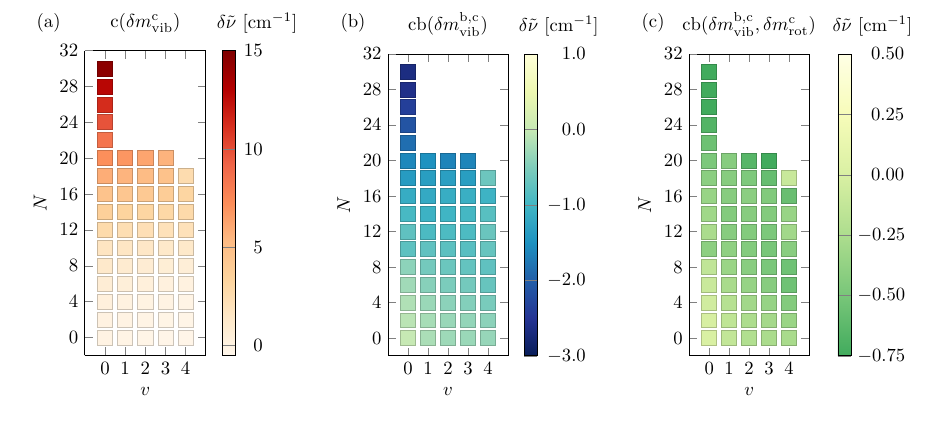} 
  \caption{%
    Rovibrational intervals for \ctSgp\ He$_2$, c$vN$-c00 measured from the c00 level (with $N=0$ and $v=0$).
    Deviation of experiment (`Exp.'~\cite{LoKeBj89}) and computation (`Comp.', this work), $\delta\tilde\nu=\tilde\nu_\text{Exp.}-\tilde\nu_\text{Comp.}$ with three different theoretical models.  
    c($\delta m^{\text{c}}_\text{vib}$): single c-state description with the vibrational mass correction; 
    cb($\delta m^{\text{b,c}}_\text{vib}$): coupled bc description with the vibrational mass corrections; 
    c($\delta m^{\text{b,c}}_\text{vib},\delta m^{\text{c}}_\text{rot}$) same as (b) with the rotational mass correction for the c state (please see also text).
    In each case, all PEC corrections were included.
    \label{fig:crovib} 
  }
\end{figure*}

%
%

\begin{table}
  \caption{%
    Example b-a transition energies, $\tilde\nu^{J_\text{b},\text{b}}_{v_\text{b}N_\text{b}}-\tilde\nu^{J_\text{a},\text{a}}_{v_\text{a}N_\text{a}}$,
    and rotational-fine-structure intervals, $\tilde f$,
    computed in this work and compared with experimental values \cite{RoBrBeBr88}. 
    The $\delta\tilde{\nu}$ and $\delta\tilde{f}$ experiment-theory deviations are also shown. 
    All values are in $\cm$. The levels with $v$ and $N$ are labelled as a$v_\text{a}N_\text{a}$ and b$v_\text{b}N_\text{n}$. The computed a-state energy levels are taken from Ref.~\cite{paper1}.
    \label{tab:brvf}
  }
  \centering
  \scalebox{0.9}{%
  \begin{tabular}{@{} r@{\ }r@{\ }r@{\ }r@{\ }l@{} r@{\ }r@{}r@{\ }l@{\ } r@{}}
  \hline\hline\\[-0.35cm]
    & &
    \multicolumn{2}{c}{Expt.~\cite{RoBrBeBr88}} &&  
    \multicolumn{2}{c}{This work} &&
    \multicolumn{1}{c}{\raisebox{-0.25cm}{$\delta\tilde{\nu}^\ast$}} &	
    \multicolumn{1}{c}{\raisebox{-0.25cm}{$\delta\tilde{f}$}} \\[-0.15cm]  
    \cline{3-4}\cline{6-7} \\[-0.30cm]
    $J_\text{b}$ & 
    $J_\text{a}$ & 
    \multicolumn{1}{c}{$\tilde\nu^{J_\text{b},\text{b}}_{v_\text{b}N_\text{b}}-\tilde\nu^{J,\text{a}}_{v_\text{a}N_\text{a}}$} &
    \multicolumn{1}{c}{$\tilde f$} && 
    \multicolumn{1}{c}{$\tilde\nu^{J_\text{b},\text{b}}_{v_\text{b}N_\text{b}}-\tilde\nu^{J_\text{a},\text{a}\ \ast}_{v_\text{a}N_\text{a}}$} &
    \multicolumn{1}{c}{$\tilde f$} &&     
    & 
    \\[0.05cm]
    \hline \\[-0.30cm]
    \multicolumn{10}{l}{Electronic-fine-structure transitions ($v_\text{a}=v_\text{b}=0$):} \\
    \multicolumn{10}{c}{$N_\text{a}=N_\text{b}=1$: b01~$\leftarrow$~a01 (Q branch):} \\
    1 & 0 & 4767.4957 &   0.4049 && 4772.4570 &   0.4108 && --4.9613 & --0.0059 \\
    1 & 1 & 4767.9006 &          && 4772.8678 &          && --4.9672 &          \\
    2 & 2 & 4767.5639 &   0.3367 && 4772.4684 &   0.3994 && --4.9045 & --0.0627$^\dagger$ \\[0.10cm]
    \multicolumn{10}{c}{$N_\text{a}=N_\text{b}=3$: b03~$\leftarrow$~a03 (Q branch):} \\
    2 & 2 & 4765.0302 &	  0.2992 && 4770.0135 &   0.2982 && --4.9833 &   0.0010 \\
    3 & 3 & 4765.3294 &	         && 4770.3117 &          && --4.9823 &         \\
    4 & 4 & 4764.9907 &	  0.3387 && 4769.9728 &   0.3389 && --4.9821 & --0.0002 \\[0.10cm]
    \multicolumn{10}{c}{$N_\text{a}=N_\text{b}=5$: b05~$\leftarrow$~a05 (Q branch):} \\
    4 & 4 & 4760.4957 &   0.2996 && 4765.5038 &   0.3001 && --5.0081 & --0.0005 \\
    5 & 5 & 4760.7953 &          && 4765.8039 &          && --5.0086 &         \\
    6 & 6 & 4760.4690 &   0.3263 && 4765.4776 &   0.3264 && --5.0086 & --0.0001 \\[0.10cm]
    \multicolumn{10}{c}{$N_\text{a}=N_\text{b}=21$: b0~21~$\leftarrow$~a0~21 (Q branch):} \\
    20 & 20 & 4658.1491 &   0.3035 && 4663.7665 &   0.3017 && --5.6174 & 0.0018 \\
    21 & 21 & 4658.4526 &          && 4664.0682 &          && --5.6156 &         \\
    22 & 22 & 4658.1491 &   0.3035 && 4663.7597 &   0.3085 && --5.6106 & --0.0050 \\[0.10cm]
    
    \hline\\[-0.30cm]    
    \multicolumn{10}{l}{Electronic-rotational-fine-structure transitions ($v_\text{a}=v_\text{b}=0$):} \\
    \multicolumn{9}{c}{$N_\text{b}=2 \leftarrow N_\text{a}=1$: b02~$\leftarrow$~a01 (R branch):} \\
    1 & 0 & 4797.0112 & --0.2578 && 4801.9858 & --0.2607 && --4.9746 & 0.0029 \\    
    2 & 1 & 4796.7534 &          && 4801.7251 &          && --4.9717 &        \\
    3 & 2 & 4796.7700 & --0.0166 && 4801.7432 & --0.0181 && --4.9732 & 0.0015 \\
    \hline\\[-0.30cm]    
    \multicolumn{10}{l}{Electronic-rotational-vibrational-fine-structure transitions:} \\
    \multicolumn{9}{c}{$v_\text{b}=1 \leftarrow v_\text{a}=0$, $N_\text{a}=N_\text{b}=3$: b13~$\leftarrow$~a03 (Q branch):} \\
    2 & 2 & 4731.9190 &   0.2915 && 4737.1109 &   0.2947 && --5.1919 & --0.0032 \\    
    3 & 3 & 4732.2105 &          && 4737.4055 &          && --5.1950 &           \\
    4 & 4 & 4731.8784 &   0.3321 && 4737.0704 &   0.3351 && --5.1920 & --0.0030 \\    
  \hline\hline\\[-0.35cm]    
  \end{tabular} 
  }
  \begin{flushleft}
  {\footnotesize
    \vspace{-0.25cm}
    $^\ast$~By further extension ($\Nb=1500$) and optimisation of the b-state fECG basis set ($\rho=2$~bohr, Table~S1), the electronic energy is already lowered (variationally) by $2.1\ \cm$ (Table~S1), which can be used as a correction, $\tilde\nu^{J,\text{b}}_{vN}-\tilde\nu^{J,\text{a}}_{vN}-2.1\ \cm$ and $\delta\tilde\nu+2.1\ \cm(\approx -2.9\ \cm)$. \\
    $^\dagger$~Perhaps a misprint in Ref.~\citenum{RoBrBeBr88} (4767.5639 vs. 4767.5039)?
  }
  \end{flushleft}
\end{table}

First of all, we aim to assess the accuracy of the `absolute' position of the b and c PECs, referenced to the more accurate a-state results \cite{paper1}.
Ref.~\citenum{FoBeCo98} informs us about measured c-a rovibronic intervals, and comparison shows that the $\text{c}v'N'-\text{a}vN=\text{c}00-\text{a}00$ electronic-vibrational energy interval differs from the experimental (effective Hamiltonian) data by $-2.0\ \cm$. The a-PEC~\cite{paper1} is converged to ca.~0.7~$\cm$ (upper bound, and with further variational optimisation, this error was already reduced to 0.3~$\cm$ in Ref.~\citenum{paper1}) at $\rho=2$~bohr. Hence, the c-a electronic excitation energy error is dominated by the variational convergence error of the c-state electronic energy near equilibrium; the c~electronic energy was estimated to be too high by $\sim 10\ \mu\Eh\sim 2\ \cm$, according to the estimated convergence of the c electronic energy (Table~S1, \som). 
Regarding the b state, comparison with the experimental (fine-structure resolved) b-a transitions~\cite{RoBrBeBr88} shows that the computational result is too large by $\approx 5\ \cm$ (Table~\ref{tab:brvf}), which is dominated by the convergence error of the b-state electronic energy. Indeed, the \btPg\ electronic energy was estimated to be $20-25\ \mu\Eh$ too high at $\rho=2$~bohr (variational upper bound), which is $\approx 5\ \cm$ and in agreement with experiment-theory deviation in the electronic excitation energy. (According to Table~S1, the b-state electronic energy obtained with the $\Nb=1000$ basis set, which was used for the PEC generation, has already been improved (lowered) by $9.5\ \mu\Eh\approx 2.1\ \cm$ using $\Nb=1500$ basis functions. This value can be used to correct the computed rovibronic b-a excitation energies in Table~\ref{tab:brvf} according to the footnote $^\ast$ of the table.) 
Both the b and c BO electronic energies can be better converged in the future (with more computing time), if necessary. 
We expect the rovibrational and fine-structure intervals to be more accurate than the electronic excitation energies, since the approximately homogeneous local error of the PEC largely cancels in the rovibrational (and fine-structure) intervals.

\begin{table}
  \caption{
    Computed c fine structure splittings, $\tilde\nu_{N\pm 1}-\tilde\nu_{N}$ in $\cm$ with the fully coupled b-c-B-C (bcBC) rovibronic model including non-adiabatic, relativistic (r) and QED couplings (Q). 
    Deviation of the ($\pm$1) fine structure components, $\delta\nu^{\pm 1}$ in MHz, from this bcBC(rQ) model obtained with 
    bc(rQ): coupled b-c model with non-adiabatic, relativistic and QED couplings; 
    bc(r): the coupled b-c model with non-adiabatic and relativistic couplings;
    and c(rQ): the single c-state model with relativistic and QED zero-field splitting. 
    \label{tab:cfine}  
  }
  \scalebox{0.9}{%
  \begin{tabular}{@{}ccc   c rr   c rr  c rr@{}}
    \hline\hline\\[-0.35cm]  
       &
       \multicolumn{2}{c}{bcBC(rQ)} &&
       \multicolumn{2}{c}{bc(rQ)} &&
       \multicolumn{2}{c}{bc(r)} &&
       \multicolumn{2}{c}{c(rQ)} \\
       \cline{2-3} \cline{5-6} \cline{8-9} \cline{11-12} \\[-0.35cm]
       $N$  & 
       \multicolumn{1}{c}{$\tilde\nu_{N-1}-\tilde\nu_{N}$} &
       \multicolumn{1}{c}{$\tilde\nu_{N+1}-\tilde\nu_{N}$} &&
       \multicolumn{1}{c}{$\delta\nu^{-1}$} &
       \multicolumn{1}{c}{$\delta\nu^{+1}$} &&
       \multicolumn{1}{c}{$\delta\nu^{-1}$} &
       \multicolumn{1}{c}{$\delta\nu^{+1}$} &&
       \multicolumn{1}{c}{$\delta\nu^{-1}$} &
       \multicolumn{1}{c}{$\delta\nu^{+1}$} \\
    \hline\\[-0.35cm]
      \multicolumn{3}{l}{c, $v=0$:} \\
        0	&		        &	0.020 234	&&		    &	--0.1	&&		&	1.3	&&		    &	0	    \\
	2	&	0.039 422	&	0.026 643	&&	--0.1	&	--0.1	&&	2.7	&	1.7	&&	--34    &	18  	\\
	4	&	0.033 102	&	0.028 839	&&	--0.1	&	--0.1	&&	2.3	&	1.8	&&	--50    &	37  	\\
	6	&	0.030 957	&	0.030 212	&&	--0.1	&	--0.1	&&	2.2	&	1.9	&&	--68    &	55  	\\
	8	&	0.029 659	&	0.031 288	&&	--0.1	&	--0.1	&&	2.1	&	1.9	&&	--85    &	74  	\\
	10	&	0.028 681	&	0.032 227	&&	--0.1	&	--0.1	&&	2.1	&	2.0	&&	--103	&	93  	\\
    \multicolumn{12}{c}{c, $v=1,2,$ and 3: \som} \\
    \multicolumn{3}{l}{c, $v=4$:} \\
	0	&		        &	0.022 045	&&		    &	--0.1	&&		&	1.5	&&		    &	0	    \\
	2	&	0.042 891	&	0.029 123	&&	--0.2	&	--0.1	&&	2.9	&	1.9	&&	--39	&	23  	\\
	4	&	0.035 942	&	0.031 597	&&	--0.2	&	--0.1	&&	2.4	&	2.0	&&	--59	&	45  	\\
	6	&	0.033 546	&	0.033 185	&&	--0.2	&	--0.1	&&	2.3	&	2.1	&&	--80	&	68  	\\
	8	&	0.032 072	&	0.034 461	&&	--0.2	&	0.0	    &&	2.2	&	2.2	&&	--102	&	91  	\\
	10	&	0.030 948	&	0.035 600	&&	--0.2	&	0.0	    &&	2.2	&	2.2	&&	--123	&	115  	\\
    \hline\hline
  \end{tabular}
  }
\end{table}

Regarding the rovibronic structure of the c state, we use the rovibrational energy intervals from the experimental work reported in Ref.~\citenum{LoKeBj89} for comparison.
Ref.~\citenum{LoKeBj89} also reports the lifetime of $v=4$ and 5 vibrationally excited states. 
In this work, we compute bound rovibrational states, and the positions of long-lived resonances (relevant for $v=4$) were first estimated using the stabilisation method~\cite{HaTa70}. Converging resonance positions and widths, necessary for $v=5$ states (and $v=4$, $N>10$), were incorporated into our rovibronic approach using complex absorbing potentials (CAP) \cite{PoCa03} (alternatively, the complex coordinate rotation technique could also be used) \cite{Mo98}. Both options require repeated computations, and CAP computations were performed with the finalised and validated rovibronic description of the bc bound-state level structure (Fig.~\ref{fig:crovib}c).

Fig.~\ref{fig:crovib} shows the deviation of experiment and our computations using three different c rovibrational (rovibronic) models. 
Fig.~\ref{fig:crovib}a summarizes the result of a single c-state model: it includes all PEC corrections and the vibrational mass correction in the vibrational kinetic energy, but using the nuclear mass for rotations (\emph{i.e.,} without any rotational mass correction, which would be singular (and undefined) due to the crossing with the b state). In this single c-state model, highly rotationally excited states exhibit large deviations from experiment, up from 10 to 15~$\cm$. 
Fig.~\ref{fig:crovib}b showcases the bc coupled model, which includes the bc non-adiabatic coupling, all PEC corrections, and the vibrational mass corrections for both the b and c states. The high-$N$ rotational angular momentum states deviate from experiment by up to $-3\ \cm$, which is (in absolute value) by a factor of 5 smaller than the single c-state result. 
Fig.~\ref{fig:crovib}c shows pilot results from the same bc coupled model, but appended with the c-state rotational mass corrections, which can be rigorously defined for the coupled bc subspace, but the evaluation of the reduced resolvent requires projecting out the active b electronic state contributions \cite{MaTe19}. 
The computation of the c rotational mass corrections assumes an auxiliary basis set of $^3\Pi_\text{g}$ symmetry (same as for b) to represent the reduced (\emph{i.e.,} without bc contributions) resolvent, Eq.~\eqref{eq:vibrot-nadmass}. Including this c-state rotational mass correction in the bc coupled computation reduces the experiment-theory deviations for high-$N$ states by an additional factor of 3, to less than 1~$\cm$, which is clearly an improvement (by an order of magnitude) over the 10-15~$\cm$ deviation of Fig.~\ref{fig:crovib}a.
At the same time, by comparison with the spectacularly accurate ($\sim 0.001~\cm$) a-state rotational intervals \cite{paper1}, we may wonder what is missing here. First, the auxiliary basis set used in the c rotational mass correction should be optimised more tightly (currently limited by technical difficulties for projecting out the b state). Furthermore, the b state also has a rotational mass correction, and the non-adiabatic mass (kinetic) energy correction has (b-c) off-diagonal terms \cite{MaTe19}, which must also be considered. All the missing technical, computational, and formal relations will be elaborated in future work.

\vspace{0.5cm}
Regarding the fine structure of the c-state rovibronic levels, Table~\ref{tab:cfine} reports fine-structure splittings for selected vibrational states. Since no well-resolved experimental data are currently available, we compare four theoretical models to gain a better understanding of the role of the different contributions. All cases include all PEC corrections and non-adiabatic couplings (in multi-state cases). The currently most complete treatment, labelled as bcBC(rQ), includes all twelve electronic-spin states of the b, c, B, and C electronic states (Tables~S6, S7) coupled by relativistic QED (rQ) interactions. The second most comprehensive model, bc(rQ), includes all nine electron-spin states of the b and c electronic states, and they are coupled by rQ couplings. The bcBC(rQ)-bc(rQ) deviation is (less) than 100~kHz, which is a surprisingly small effect, if we note that the b-B, the b-C, and c-B relativistic (QED) couplings (Fig.~\ref{fig:allcurves}) are comparable to those of the b-b and b-c states. The important difference is that the b-c states are coupled non-adiabatically, whereas the triplet-singlet (bc-BC) states are not (for their different electron spins). The non-adiabatic coupling seems to `enhance' the effect of the relativistic couplings. It is interesting to compare the bc(rQ) and the bc(r) results, the latter including only the relativistic couplings but neglecting the QED corrections of the couplings (from the anomalous magnetic moment of the electron), which has a `large', 2-3~MHz effect (similar to the a-state fine structure \cite{paper1}). Interestingly, the QED effect on the bc magnetic couplings is larger than any magnetic couplings with the BC states. Finally, a single c-state computation (by neglecting any non-adiabatic or relativistic coupling with the b states), but still 
including the relativistic QED zero-field splitting (within the \ctSgp electron spin subspace) has a significant, 30-120~MHz, effect compared to our currently most comprehensive bcBC(rQ) [or almost equally good bc(rQ)] model.

Finally, we report b-a electronic-rotational-vibrational-fine-structure transition energies (Table~\ref{tab:brvf}) obtained from bc(rQ) coupled computations. For a direct, line-by-line comparison with the experimental b-a transitions, we take the (highly accurate) a-state rovibrational-fine-structure energy levels from Ref.~\citenum{paper1}. The couplings with the B-C PECs were not included in these results, because they have a negligibly small effect (see also the c-state fine structure analysis, Table~\ref{tab:cfine}). 
By inspecting Table~\ref{tab:brvf}, we can assess the accuracy of the b representation at different scales: electronic, vibrational, rotational, and electron spin (fine structure). As pointed out earlier in this work, the b PEC is by $\sim 5\ \cm$ too high in energy (out of which $2.1~\cm$ is already confirmed in a larger-basis, currently single $\rho=2$~bohr point computation, Table~S1). The first vibrational fundamental is accurate to 0.2~$\cm$. High-$N$ (up to $N=21$ is available from experiment) rotationally excited states are less accurately described, transitions to $N=21$ b rotational states are by ca. 0.6~$\cm$
less accurate than to $N=1$ (again, we emphasise that the a-state rovibrational levels are much more accurate \cite{paper1}, so their theoretical-computational uncertainty is negligible in the present analysis).
For a more accurate computation of the b-state rotational excitations, in a bc coupled model, it will be necessary to compute the rotational mass correction of the b state (and of the bc off-diagonal terms); nevertheless, the explicit coupling with the c-state ensures that the deviation is not larger. For the b-state rotational mass correction, auxiliary basis sets of $^3\Sigma_\text{g}^+$ and $^3\Delta_\text{g}$ symmetries must be optimised \cite{MaFe22nad}, but the c-state contribution must be projected out, since it is already explicitly coupled with the b state. So, the largest rotational contribution is already accounted for by the explicit bc non-adiabatic coupling.
The $\tilde{f}$ fine-structure splittings are in excellent agreement with the experimental data; their deviation from the experimental value is $\delta\tilde{f}\approx 0.001-0.003$~$\cm$, which matches the experimental resolution of Ref.~\citenum{RoBrBeBr88}.

In summary, we have developed a high-precision rovibronic-relativistic-QED model of the b-c electronic states of the triplet helium dimer. The computed results significantly improve upon earlier work: they account for the electronic, vibrational, rotational and fine-structure components of the rovibronic spectrum in quantitative agreement with available experimental data. As a result, we believe that this rovibronic-electron-spin \emph{ab initio} model can aid ongoing and future experimental work on this simple diatomic system, which exhibits rich magnetic properties. 
During the introduction of the formalism and the analysis of the computed results, several directions for future improvements were identified, including better convergence of the PECs, non-adiabatic relativistic couplings, b-state rotational mass corrections, spin-rotation coupling computation, and consideration of further electronic states. 
Priority will be given to the most experimentally useful directions. 
In the meantime, we are also working on a control program that organises the several separate or consecutive QUANTEN computing jobs for converging the electronic energy, computing the PEC and all corrections and couplings at the PEC points. Thereby, within the theoretical framework presented in this paper, (even) more accurate and larger-scale computations can be conducted in the future.

\section*{Supplementary Material}
\noindent %
The Supplementary Material includes %
(a) further theoretical, computational and convergence details; 
(b) datapoints for the PECs, correction and coupling curves are deposited, and their applications are demonstrated in Wolfram Mathematica notebook files; 
(c) rovibrational and fine structure energy lists for the b and the c electronic states.

\section*{Acknowledgement}
\noindent %
We thank the European Research Council (Grant No. 851421) and the Momentum Programme of the Hungarian Academy of Sciences (LP2024-15/2024) for their financial support. 
PJ acknowledges the support of the János Bolyai Research Scholarship of the Hungarian Academy of Sciences (BO/285/22).
We thank DKF 
for access to the Komondor HPC facility.
This paper is dedicated to John Stanton's memory. We thank John for drawing our attention to the double- and triple-augmented Dunning basis sets.

%

\input{bcHe2_som_inc}

\end{document}

%% file: bcHe2_som_inc.tex
\clearpage
\begin{center}
{\large
\textbf{Supplemental Material}
}\\[0.25cm]
{\large
\textbf{High-Precision Quantum Dynamics of He$_2$ (b~$^3\Pi_\text{g}$-c~$^3\Sigma_\text{g}^+$) including Non-adiabatic, Relativistic and QED Corrections and Couplings}
} \\[0.5cm]

Balázs Rácsai,$^1$ Péter Jeszenszki,$^1$ Ádám Margócsy,$^1$ Edit Mátyus$^{1,\ast}$ \\
\emph{$^1$~MTA–ELTE `Momentum' Molecular Quantum electro-Dynamics Research Group,
Institute of Chemistry, Eötvös Loránd University, Pázmány Péter sétány 1/A, Budapest, H-1117, Hungary} \\
$^\ast$ edit.matyus@ttk.elte.hu 
~\\[0.15cm]
(Dated: 2 September 2025)
\end{center}

\setcounter{section}{0}
\renewcommand{\thesection}{S\arabic{section}}
\setcounter{subsection}{0}
\renewcommand{\thesubsection}{S\arabic{section}.\arabic{subsection}}

\setcounter{equation}{0}
\renewcommand{\theequation}{S\arabic{equation}}

\setcounter{table}{0}
\renewcommand{\thetable}{S\arabic{table}}

\setcounter{figure}{0}
\renewcommand{\thefigure}{S\arabic{figure}}

~\\
The numerical values of the physical constants and conversion factors were taken from CODATA 2022~\cite{codata22}; they are
$\alpha = 0.007\ 297\ 352\ 564\ 3$, 
$1\ \Eh = 219\ 474.631\ 363\ 14\, (hc)\text{cm}^{-1}$,
and the mass of the helium nucleus, $M(^4\text{He}) = 7\ 294.299\ 541\ 71\ m_\text{e}$, 
$\Eh = 6.579683920 4999\cdot10^{+9}\ (h \, \text{MHz})$.  \\

\section{Leading-order relativistic and QED correction terms}
In this section, we provide a short summary of the expressions used; further implementation details are reported for the spin-independent terms in Refs.~\citenum{JeIrFeMa22,RaFeMaMa24} and for the spin-dependent terms in Ref.~\citenum{paper3}.
The electronic energy is approximated as the sum of the non-relativistic energy, $E^{(0)}_\text{x}$ (with wave function, $\varphi^{(0)}_\text{x}$; solutions of electronic Schrödinger equation), and relativistic and quantum electrodynamics (QED) corrections written as powers of the $\alpha$ fine-structure constant,
\begin{align}
  E_\text{x} 
  = 
  E^{(0)}_\text{x} 
  + \alpha^2 \langle \varphi^{(0)}_\text{x} | \hat{H}^{(2)}_\text{BP,sn} | \varphi^{(0)}_\text{x} \rangle  
  + \alpha^2 \langle \varphi^{(0)}_\text{x} | \hat{H}^{(2)}_\text{BP,sd} | \varphi^{(0)}_\text{x} \rangle    
  + \alpha^3 E^{(3)}_\text{x}  
   + \alpha^3 \langle \varphi^{(0)}_\text{x} | \hat{H}^{(3)}_\text{sd} | \varphi^{(0)}_\text{x} \rangle  
\end{align}
For convenience, we write the relativistic and QED corrections as the sum of the spin-independent (sn) centroid correction and spin-dependent (sd) terms.
The electronic states considered in this work have different spatial symmetries; therefore, we do not require any off-diagonal centroid corrections. However, the spin-dependent terms introduce off-diagonal electronic contributions, which we compute and utilise for predicting the fine-structure splitting of the rovibronic states. In this work, the electronic states with `x': b, c, B, and C are considered.

The spin-independent relativistic correction is obtained as the expectation value with the non-relativistic wave function for the Breit-Pauli (BP) Hamiltonian,
\begin{align}
  \hH^{(2)}_{\text{BP,sn}} 
  =
  \hH_\MV + \hH_\Done +  
  \hH_\OO + \hH_\Dtwo + \hH_\text{SS,c} \ 
  \label{eq:Hbp}
\end{align}
with the mass-velocity, the one-electron Darwin, the two-electron Darwin and the orbit-orbit terms, and the Fermi contact part of 
of the spin-spin interaction
\begin{align}
  \hH_\MV
  &=
  -\frac{1}{8} 
  \sum_{i=1}^{\nel} (\hbp_{i}^{2})^2 \ ,
  \\
  \hH_\Done
  &=
  \frac{\pi}{2}\hat{\delta}_1 \ \ , \ \ 
  \hat{\delta}_1=\sum_{i=1}^{\nel}
  \sum_{A=1}^{\nnuc}
    Z_A\delta(\br_{iA}) \ ,
  \\
  \hH_{\OO}
  &=
  -\frac{1}{2}
  \sum_{i=1}^{\nel}\sum_{j=i+1}^{\nel}
    \left[%
      \frac{1}{r_{ij}}\hbp_{i}\hbp_{j} + \frac{1}{r_{ij}^{3}}(\br_{ij}(\br_{ij}\hbp_{j})\hbp_{i})
    \right] \ ,
  \\
  \hH_{\Dtwo}
  &=
  -\pi\hat{\delta}_2 \ \ , \ \ 
  \hat{\delta}_2=
  \sum_{i=1}^{\nel}\sum_{j=i+1}^{\nel}
    \delta(\br_{ij}) \ ,
  \\
  \hat{H}_\text{SS,c} 
  &=  
  -\frac{8\pi}{3} 
  \sum_{i=1}^{\nel}\sum_{j=i+1}^{\nel}
    \hbs_i \hbs_j \delta(\br_{ij}) \ ,
\end{align}
respectively, and $\hat{\boldsymbol{s}}_i=I(1)\otimes ...\otimes \frac{1}{2}\boldsymbol{\sigma}(i)\otimes...\otimes I(\nel)$.

The leading-order QED correction to the centroid energy is
\begin{align}
  E^{(3)}_\text{x}
  &=
  \frac{4}{3}
  \left[%
    \frac{19}{30}-2\ln\alpha-\ln(k_0)_\text{x} 
  \right]
    \langle \varphi^{(0)}_\text{x} | 
      \sum_{i=1}^{\nel} \sum_{A=1}^{\nnuc}Z_A\delta(\br_{iA}) 
    |\varphi^{(0)}_\text{x} \rangle
  \nonumber \\ 
  &+
  \left[%
    \frac{164}{15}+\frac{14}{3}\ln\alpha 
  \right]
  \langle \varphi^{(0)}_\text{x} | 
    \sum_{i=1}^{\nel}\sum_{j=i+1}^{\nel} \delta(\br_{ij}) 
  | \varphi^{(0)}_\text{x} \rangle
  -\frac{7}{6 \pi}
  \langle \varphi^{(0)}_\text{x} | 
    \sum_{i=1}^{\nel}\sum_{j=i+1}^{\nel}{\cal{P}}\left(1/r_{ij}^3 \right) 
  | \varphi^{(0)}_\text{x} \rangle
  \ .
  \label{eq:E3x}
\end{align}
with the Bethe logarithm,
\begin{align}
  \ln(k_0)_\text{x}
  =
  \frac{%
    \left\langle %
      \varphi^{(0)}_\text{x} |
      \hat{\boldsymbol{P}}_\text{e}(\hat{H}_\text{e}-E_{\text{x}})
      \ln[2|\hat{H}_\text{e}-E_{\text{x}}|]
      \hat{\boldsymbol{P}}_\text{e} |
      \varphi^{(0)}_\text{x}
    \right\rangle}{2\pi \langle \varphi_\text{x}^{(0)} | \hat{\delta}_1 | \varphi_\text{x}^{(0)} \rangle} \; ,
 \label{eq:BLterm}
\end{align}
where $\hat{H}_\text{e}$ and $E^{(0)}_\text{x},\varphi^{(0)}_\text{x}$ are the non-relativistic Hamiltonian and the electronic energy and wave function. $\hat{\boldsymbol{P}}_\text{e}=\sum_{i=1}^{\ne} \hat{\boldsymbol{p}}_i$ is the total momentum of electrons.
The Araki-Sucher distribution (with the Euler-Mascheroni constant, $\gamma$) is
\begin{align}
  {\cal{P}}\left(\frac{1}{r_{ij}^3}\right)=\lim_{\epsilon\rightarrow0^+}\left[\frac{\Theta(r_{ij}-\epsilon)}{r_{ij}^3}+4\pi [\gamma+\ln(\epsilon)]\delta(\boldsymbol{r}_{ij})\right] \ .  
 \label{ASterm}
\end{align}
The Bethe logarithm is approximated by the (single-electron) ion-core value \cite{FeKoMa20,FeMa22bethe} for all electronic states considered in this work (see also Ref.~\cite{paper1}). 
The AS term is computed for the c state with the integral transform technique \cite{PaCeKo05} over the $\rho\in[1,10]$~bohr internuclear distances interval, and it is used (as an approximation) for the b, B, and C states.
The higher-order QED corrections and finite-nuclear-size effects are neglected in this work.

The leading-order relativistic spin-dependent corrections in the Breit-Pauli Hamiltonian are
\begin{align}
  \hat{H}^{(2)}_\text{BP,sd}
  =
  \hat{H}_\text{SO}+\hat{H}_\text{SOO}+\hat{H}  _{\text{SOO}^{'}}+\hat{H}_\text{SS,dp}
  \label{eq:Hsd2}
\end{align}
\noindent
with 
the one-electron spin-orbit,
the spin-own-orbit, 
the spin-other-orbit, and 
the magnetic dipole part of the spin-spin interaction,
\begin{align}
  \hat{H}_{\text{SO}}
  &=
  \frac{1}{2}\sum_{i=1}^{\nel} %
  \sum_{A=1}^{\nnuc}\frac{Z_A}{r_{iA}^3}\hat{\boldsymbol{s}}_i\cdot \hat{\boldsymbol{l}}_{i,A} \ ,
  \nonumber \\
  \hat{H}_{\text{SOO}}
  &=
  -\frac{1}{2}\sum_{i=1}^{\nel}\sum_{j=i+1}^{\nel} %
  \frac{1}{r_{ij}^3}\left[\hat{\boldsymbol{s}}_i\cdot \hat{\boldsymbol{l}}_{i,j}+\hat{\boldsymbol{s}}_j\cdot \hat{\boldsymbol{l}}_{j,i}\right] \ ,
  \nonumber \\
  \hat{H}_{\text{SOO}^{'}}
  &=-
  \sum_{i=1}^{\nel}\sum_{j=i+1}^{\nel} %
  \frac{1}{r_{ij}^3}\left[\hat{\boldsymbol{s}}_i\cdot\hat{\boldsymbol{l}}_{j,i}+\hat{\boldsymbol{s}}_j\cdot\hat{\boldsymbol{l}}_{i,j}\right] \ ,
  \nonumber \\
  \hat{H}_\mathrm{SS,dp}
  &=
  \sum_{i=1}^{\nel}\sum_{j=i+1}^{\nel} %
    \left[%
      \frac{\hbs_{i} \hbs _{j}}{r_{ij}^3}
      -3\frac{\left(\hbs_{i}\bos{r}_{ij}\right)\left(\hbs_{j}\bos{r}_{ij}\right)}{r_{ij}^5}
  \right] \ ,
\end{align}
respectively, with $\hat{\boldsymbol{l}}_{a,b}=\boldsymbol{r}_{ab}\times \hat{\boldsymbol{p}}_a$.

The leading-order QED corrections to the spin-dependent couplings are due to the QED correction to the free-electron $g$ factor, 
$g=2\left[1+\frac{\alpha}{2\pi}+{\cal{O}}(\alpha^2)\right]$, 
\begin{align}
  \alpha^3 \hat{H}^{(3)}_\text{sd}
  &=
  \alpha^2
  \left[%
    \frac{\alpha}{\pi} %
    (\hat{H}_\text{SO}+\hat{H}_\text{SOO})
    +
    \frac{\alpha}{2\pi} \hat{H}_{\text{SOO}^{'}}
    +
    \frac{\alpha}{\pi} \hat{H}_\text{SS,dp}
  \right] \; .
\end{align}

\section{Convergence tables}
The convergence of the \btPg\ and \ctSgp\ electronic energy for $\rho=2$~bohr internuclear distance is shown in Table~\ref{tab:bcE}. 
The table also reviews literature data and standard quantum chemistry computations (full configuration interaction, FCI) performed by us using the Molpro program package \cite{molpro1}. With standard approaches and basis sets near the 1~m$\Eh$\  ($10^{-3}\ \Eh$) convergence of the electronic energy could be reached only with the (doubly and) triply augmented correlation consistent basis sets.
Sub-m$\Eh$\ convergence was easily reached with variationally optimised fECGs (already with $\Nb=200$ optimised fECGs, which amounts to a few hours of computation on a PC).
We also report the electronic energy for the \dtSup\ and \etPg\ (preliminary computations, providing upper bounds to the exact value) and compare it with available literature data for potential future relevance.
In addition, small-scale fECG computations were also performed for the \BsPg\ and \CsSgp\ electronic states, for their significant spin-dependent relativistic couplings with the b and c states. 
The computational details and convergence for B and C are discussed in Ref.~\cite{paper3}.

After the $\rho=2$~bohr variational generation and optimisation of fECG basis sets, the PECs were generated by performing small consecutive displacements of $\Delta\rho=\pm 0.1$~bohr followed by the rescaling of the fECG centres and full reoptimization of the non-linear basis parameters \cite{CeRy95,FeMa19HH,FeKoMa20,FeMa22h3}. The consecutive rescaling and reoptimisation of the fECG parameterisation ensures that the local relative error of the PEC is smaller than the absolute convergence error of the total energy.

The convergence of the relativistic corrections for the b and c states is presented in Tables~\ref{tab:brel} and \ref{tab:crel}, respectively. In the rovibronic computations, we use the regularised values obtained with the numerical Drachmanization technique \cite{Dr81,RaFeMaMa24}, which we found to be more robust than the integral transformation (IT) technique \cite{JeIrFeMa22,PaCeKo05}.
The relativistic corrections for the B and C states were also computed with the numerical Drachmanization approach. We only note that the C state is the first excited state in the $^1\Sigma_\text{g}$ symmetry block, and this fact must be considered (use of the corresponding energy eigenvalue) in the Drachmanized expressions \cite{Dr81,RaFeMaMa24}.
The leading-order QED corrections are calculated using the Dirac-delta expectation values already computed for the relativistic corrections. We used the ion-core approximation for the Bethe logarithm, computed in Ref.~\cite{FeKoMa20}, and tested its accuracy for the \atSup\ state in Ref.~\cite{paper1}. 
We computed the Araki-Sucher (AS) term for the c state in this work using the IT technique. The computation was repeated at all PEC points. The AS term provides a small correction to the energy (even smaller for relative rotational and vibrational intervals). We noticed that its value was very similar for the c and for the a states Ref.~\cite{paper1}. So, the b, B, and C leading-order QED corrections were computed using the AS correction curve taken from the c state (as a good approximation, considering the PEC error, etc.).

The convergence of the spin-dependent relativistic (and QED) couplings is discussed in detail in Ref.~\cite{paper3}. 

Table~\ref{tab:elang} shows convergence tests for the electronic angular momentum matrix elements, for the non-adiabatic and DBOC contributions. 
Table~\ref{tab:eldip} shows the convergence of the electric dipole transition moment.

In the rovibronic computations, the $\Nb=1000$ basis set of the b state was used for the PEC, DBOC and non-adiabatic couplings over the $\rho\in[1,10]$~bohr interval, whereas the relativistic corrections and couplings were computed with the $\Nb=750$ basis set also available for $\rho\in[1,10]$~bohr.
Regarding the c state, the PEC, all corrections curves and couplings were computed with the $\Nb=1500$ basis set of Table~\ref{tab:bcE}.

\begin{table}
  \caption{%
    Electronic energy, in $\Eh$, for electronically excited states of He$_2$ ($\rho=2$~bohr): 
    \btPg, \ctSgp, and pilot computations for \dtSup, and \etPg.
    \label{tab:bcE}
  }
  \begin{tabular}{@{}lll ll l@{}}
    \hline\hline\\[-0.35cm]
      & 
      $U[\text{\btPg}]$ & 
      $U[\text{\ctSgp}]$ & 
      $U[\text{\dtSup}]$ & 
      $U[\text{\etPg}]$ &       
      \\
    \hline\\[-0.35cm]
    MRCI \cite{Ya89,ChJeYaLe89}$^\text{a,b}$
                                      & --5.123 5   & --5.096 7 &  &&  \\
    MRCI \cite{BjMiPaRo98}$^\text{b}$ & --5.128 449 & \multicolumn{1}{c}{--}        &  &&  \\
    MRCI \cite{XuLuZhGuSh24}$^\text{c}$ & \multicolumn{1}{c}{--} & --5.091 &  &&  \\
    \hline \\[-0.35cm]
    \multicolumn{4}{l}{FCI with Molpro \cite{molpro1} [this work]:} \\    
    FCI/aug-cc-pVDZ    & --4.95    & --5.04      & --4.70     & --4.25    & \\
    FCI/aug-cc-pVTZ    & --5.023   & --5.081     & --4.81     & --4.38    & \\
    FCI/aug-cc-pVQZ    & --5.060   & --5.088     & --4.87     & --4.46    & \\
    FCI/aug-cc-pV5Z    & --5.079   & --5.091     & --4.91     & --4.51    & \\[0.15cm]
    FCI/daug-cc-pVDZ   & --5.10    & --5.076     & --5.001    & --4.83    & \\
    FCI/daug-cc-pVTZ   & --5.12    & --5.095     & --5.043 3  & --4.88    & \\
    FCI/daug-cc-pVQZ   & --5.126   & --5.098 6   & --5.047 7  & --4.925   & \\
    FCI/daug-cc-pV5Z   & --5.127 6 & --5.099 76  & --5.048 3  & --4.953   & \\[0.15cm]
    FCI/taug-cc-pVDZ   & --5.10    & --5.076     & --5.032    & --5.027   & \\
    FCI/taug-cc-pVTZ   & --5.124   & --5.095     & --5.051 66 & --5.043 5 & \\
    FCI/taug-cc-pVQZ   & --5.127 6 & --5.098 9   & --5.055 42 & --5.047 2 & \\
    FCI/taug-cc-pV5Z   & --5.128 6 & --5.099 8   & --5.056 69 & --5.049 4 &\\    
    \hline\\[-0.35cm]
    $R$ matrix \cite{EpMoChTe24}$^\ast$ 
                       & --5.124 7 & --5.096 4    & --5.055 19 & --5.051 82 & \\    
    \hline\hline\\[-0.35cm]
    \multicolumn{4}{l}{Variational ECG (QUANTEN) [this work]:$^\text{d}$} \\
    20	    & --5.11 	        & --5.07  \\
    50	    & --5.124 	        & --5.093  \\
    100	    & --5.127 9	        & --5.096 6 \\
    200	    & --5.128 8      	& --5.099 6  \\
    500	    & --5.129 25    	& --5.100 32  \\
    750  	& --5.129 307$^\text{e}$ & --5.100 426 	\\
    1000	& --5.129 329 6$^\text{e}$  	& --5.100 464 7	\\
    1250	& --5.129 335 71    & --5.100 475 59 
    &  
    &
    --5.052 834 8$^\text{h}$ & \\
    
    1500    & 
    --5.129 339 08
            & 
    --5.100 481 76$^\text{f,g}$ & --5.058 021 9$^\text{i}$ & \\
        & 
        & 
    --5.100 482 65 & & \\
    \hline\hline\\[-0.40cm]           
  \end{tabular}
  \begin{flushleft}
    $^\text{a}$ Ref.~\citenum{EpMoChTe24} reported relative values with respect to the \atSup\ state energy from Ref.~\citenum{Ya89}.  \\
    $^\text{b}$ Special basis set with diffuse functions. \\
    $^\text{c}$ Extrapolated energy using aug-cc-pVnZ basis sets. \\
    $^\text{d}$ The number of symmetry-adapted ECG functions ($\Nb$) is listed in the first column. \\
    $^\text{e}$ Basis sets used for PEC generation in this work. \\
    $^\text{f}$ Obtained from the \atSup\ parameterisation \cite{paper1}, after changing the symmetry, we have carried out several Powell refinement cycles of the entire basis set. \\
    $^\text{g}$ This basis set was used for the \ctSgp\ PEC generation in this work. \\
    $^\text{h}$ Obtained from further optimization of the \atSup\ $\Nb=1500$ (Ref.~\citenum{paper1}) basis set for the second electronic state. \\
    $^\text{i}$ Obtained from further optimization of the \btPg\ $\Nb=1250$ basis set for the second electronic state. \\
  \end{flushleft}
\end{table}

\begin{table}[h!]
  \caption{%
    \btPg~He$_2$ ($\rho=2$~bohr): 
    Convergence of the relativistic corrections (in atomic units, to be multiplied by $\alpha^2$), without regularisation (direct evaluation) and with the numerical Drachmanization approach \cite{RaFeMaMa24}.
    \label{tab:brel}
  }
    \centering
    \begin{tabular}{@{}r@{\ \ \ \ } c@{\ \ \ \ } c@{\ \ \ \ } c@{\ \ \ \ }  c@{\ \ \ \ } c@{}}
    \hline\hline\\[-0.35cm]
       $\Nb$ & 
       $\langle \hH_\MV \rangle$ &
       $\langle \hH_\Done \rangle$ &
       $\langle \hH_\Dtwo \rangle$ & 
       $\langle \hH_\OO \rangle$$^\text{a}$ &   
       $\langle \hH^{(2)}_\text{BP,sn} \rangle$ \\
       \hline \\[-0.35cm]
       \multicolumn{6}{l}{Direct evaluation:} \\    
       20	 &	--21.406 904	&	17.462 862	&	--0.438 322	&	--0.042 397	&	--3.548 117	\\
       50	 &	--22.165 453	&	18.118 236	&	--0.419 458	&	--0.033 837	&	--3.661 596	\\
       100	 &	--22.850 082	&	18.743 318	&	--0.393 510	&	--0.031 681	&	--3.744 936	\\
       200	 &	--23.001 121	&	18.876 257	&	--0.389 795	&	--0.030 917	&	--3.765 986	\\
       500	 &	--23.106 521	&	18.976 478	&	--0.381 500	&	--0.030 285	&	--3.778 828	\\
       750	 &	--23.359 460	&	19.208 785	&	--0.381 183	&	--0.030 207	&	--3.799 699	\\
       1000 & 	--23.374 099	&	19.224 027	&	--0.378 770	&	--0.030 109	&	--3.801 411	\\
       \hline\\[-0.35cm]
       \multicolumn{6}{l}{Numerical Drachmann regularization:} \\
       20	 &	--23.378 789	&	19.342 105	&	--0.356 743	&	--	&	--3.722 338	\\
       50	 &	--23.603 855	&	19.479 360	&	--0.363 593	&	--	&	--3.794 739	\\
       100	 &	--23.709 769	&	19.551 920	&	--0.368 639	&	--	&	--3.820 891	\\
       200	 &	--23.740 855	&	19.570 023	&	--0.369 879	&	--	&	--3.831 870	\\
       500	 &	--23.753 778	&	19.580 085	&	--0.370 651	&	--	&	--3.833 327	\\
       750	 &	--23.769 189	&	19.589 906	&	--0.370 724	&	--	&	--3.838 767	\\
       1000 &	--23.769 922	&	19.591 181	&	--0.370 840	&	--	&	--3.838 009	\\       
    \hline\hline \\[-0.50cm]
    \end{tabular}
    \begin{center}
      $^\text{a}$ The $\hH_\OO$ term is from direct computation; no regularisation is needed. \\
    \end{center}
\end{table}

\begin{table}[h!]
  \caption{%
    \ctSgp~He$_2$ ($\rho=2$~bohr): 
    Convergence of the relativistic corrections (in atomic units), without regularisation, with the integral transformation (IT) technique \cite{PaCeKo05,JeIrFeMa22}, and with the numerical Drachmanization approach \cite{RaFeMaMa24}.
    \label{tab:crel}
  }
    \centering
    \begin{tabular}{@{}r@{\ \ \ \ } c@{\ \ \ \ } c@{\ \ \ \ } c@{\ \ \ \ }  c@{\ \ \ \ } c@{}}    
    \hline\hline\\[-0.35cm]
       $\Nb$ & 
       $\langle \hH_\MV \rangle$ &
       $\langle \hH_\Done \rangle$ &
       $\langle \hH_\Dtwo \rangle$ & 
       $\langle \hH_\OO \rangle$$^\text{a}$ &   
       $\langle \hH^{(2)}_\text{BP,sn} \rangle$ \\
       \hline\\[-0.35cm]
       \multicolumn{6}{l}{Direct evaluation:} \\    
       20	 &	--20.671 326	&	16.780 649	&	--0.480 001	&	--0.068 849	&	--3.479 525	\\
       50	 &	--22.289 192	&	18.196 139	&	--0.446 362	&	--0.081 569	&	--3.728 260	\\
       100	 &	--23.320 502	&	19.100 363	&	--0.411 218	&	--0.081 292	&	--3.890 214	\\
       200	 &	--23.562 205	&	19.336 949	&	--0.395 690	&	--0.079 879	&	--3.909 445	\\
       500	 &	--23.849 102	&	19.595 722	&	--0.389 046	&	--0.079 479	&	--3.943 812	\\
       750	 &	--23.894 195	&	19.644 650	&	--0.384 356	&	--0.079 314	&	--3.944 503	\\
       1000 &	--23.931 416	&	19.683 266	&	--0.382 466	&	--0.079 248	&	--3.944 932	\\
       1500 &	--24.085 349	&	19.828 320	&	--0.378 554	&	--0.079 177	&	--3.957 653	\\
       \hline\\[-0.35cm]
       \multicolumn{6}{l}{IT regularization:} \\ 
       20	 &	--21.139 961	&	16.946 048	&	--0.478 833	&	--	&	--3.783 930	\\
       50	 &	--22.697 331	&	18.375 481	&	--0.445 276	&	--	&	--3.958 143	\\
       100	 &	--23.764 766	&	19.293 970	&	--0.410 217	&	--	&	--4.141 872	\\
       200	 &	--24.040 099	&	19.528 964	&	--0.394 727	&	--	&	--4.196 287	\\
       500	 &	--24.131 601	&	19.772 173	&	--0.388 099	&	--	&	--4.050 808	\\
       750	 &	--24.150 013	&	19.812 694	&	--0.383 420	&	--	&	--4.033 213	\\
       1000 &	--24.131 881	&	19.838 383	&	--0.381 531	&	--	&	--3.991 215	\\
       1500 &	--24.136 822	&	19.876 211	&	--0.377 673	&	--	&	--3.962 115	\\       
       \hline\\[-0.35cm]
       \multicolumn{6}{l}{Numerical Drachamnn regularization:} \\
       20	 &	--23.381 989	&	19.362 670	&	--0.340 000	&	--	&	--3.748 169	\\
       50	 &	--23.888 333	&	19.701 827	&	--0.364 112	&	--	&	--3.903 963	\\
       100	 &	--24.061 962	&	19.823 630	&	--0.371 900	&	--	&	--3.947 724	\\
       200	 &	--24.108 199	&	19.856 015	&	--0.374 515	&	--	&	--3.957 549	\\
       500	 &	--24.132 984	&	19.872 447	&	--0.375 675	&	--	&	--3.964 342	\\
       750	 &	--24.136 197	&	19.875 735	&	--0.376 032	&	--	&	--3.963 744	\\
       1000 &	--24.137 972	&	19.877 366	&	--0.376 132	&	--	&	--3.963 721	\\
       1500 &	--24.139 190	&	19.879 490	&	--0.376 230	&	--	&	--3.962 647	\\       
    \hline\hline \\[-0.50cm]
    \end{tabular}
    \begin{center}
      $^\text{a}$ The $\hH_\OO$ term is from direct computation, no regularization is needed. \\
    \end{center}
\end{table}

\begin{table}[]
  \caption{%
    Convergence of the electronic angular momentum matrix elements with the basis size of the b and c states, $\Nb^\text{b}$ and $\Nb^\text{c}$, respectively. The nuclei are on the $z$ axis, the bra and the ket states have the same spin projection.
    \label{tab:elang}
  }
    \centering
    \begin{tabular}{@{}l ll lll l@{}}
    \hline\hline \\[-0.35cm]
      $N_\text{b}^\text{b}$ & $N_\text{b}^\text{c}$ &
      \multicolumn{1}{c}{$|\langle \bel^{x}|\hL_{x}|\cel^{0} \rangle|$} &
      \multicolumn{1}{c}{$\langle \bel^{x}|\hL^2_{x}+\hL^2_{y}|\bel^{x} \rangle$} &
      \multicolumn{1}{c}{$\langle \cel^{0}|\hL^2_{x}+\hL^2_{y}|\cel^{0} \rangle$} \\
    \hline \\[-0.35cm]
      \multicolumn{2}{l}{Ref.~\cite{Ya89}}
                & 0.863 897 &         &           \\
    \hline \\[-0.35cm]
     300	&  300 & 0.850 10 & 7.033 28 & 8.471 71 \\
     750	& 1000 & 0.850 73 & 7.036 46 & 8.468 97 \\ 
    1000 & 1500	& 0.850 79 & 7.036 65 & 8.469 06 \\
    \hline\hline      
   \end{tabular}    
\end{table}
\begin{table}[]
  \caption{%
    Convergence of the 
    electronic dipole transition moments
    with the basis size of the b and c states, $\Nb^\text{b}$ and $\Nb^\text{c}$, respectively. The nuclei are on the $z$ axis, the bra and the ket states have the same spin projection. The basis set for the a state, with $\Nb^\text{a}=1500$, is taken from Ref.~\citenum{paper1}.
    \label{tab:eldip}
  }
    \begin{tabular}{@{}l l@{\ \ \ \ \ \ \ \ }ll@{}}
    \hline\hline \\[-0.35cm]
      $N_\text{b}$ & 
      \multicolumn{1}{c}{$\langle \ael^{0}|\mu_x|\bel^{x} \rangle$} &
      $N_\text{c}$ &
      \multicolumn{1}{c}{$\langle \ael^{0}|\mu_x|\cel^{0} \rangle$}       
      \\
    \hline \\[-0.35cm]
      Ref.~\cite{Ya89} 
           & 2.674 318 &  & 3.024 279 \\
    \hline \\[-0.35cm]
     300	& 2.603 66 &  300 & 2.995 13 \\
     750	& 2.606 86 & 1000 & 2.994 12 \\
    1000 & 2.607 03 & 1500 & 2.994 08 \\
    \hline\hline      
   \end{tabular}       
\end{table}

\begin{table}
  \caption{
    More comprehensive version of Table~II: Computed c fine structure splittings, $\tilde\nu_{N\pm 1}-\tilde\nu_{N}$ in $\cm$ with the fully coupled b-c-B-C (bcBC) rovibronic model including non-adiabatic, relativistic (r) and QED couplings (Q). 
    Deviation of the ($\pm$1) fine structure components, $\delta\nu^{\pm 1}$ in MHz, from this bcBC(rQ) model obtained with 
    bc(rQ): coupled b-c model with non-adiabatic, relativistic and QED couplings; 
    bc(r): the coupled b-c model with non-adiabatic and relativistic couplings;
    and c(rQ): the single c-state model with relativistic and QED zero-field splitting. 
    \label{tab:cfine}  
  }
  \scalebox{0.9}{%
  \begin{tabular}{@{}ccc   c rr   c rr  c rr@{}}
    \hline\hline\\[-0.35cm]  
       &
       \multicolumn{2}{c}{bcBC(rQ)} &&
       \multicolumn{2}{c}{bc(rQ)} &&
       \multicolumn{2}{c}{bc(r)} &&
       \multicolumn{2}{c}{c(rQ)} \\
       \cline{2-3} \cline{5-6} \cline{8-9} \cline{11-12} \\[-0.35cm]
       $N$  & 
       \multicolumn{1}{c}{$\tilde\nu_{N-1}-\tilde\nu_{N}$} &
       \multicolumn{1}{c}{$\tilde\nu_{N+1}-\tilde\nu_{N}$} &&
       \multicolumn{1}{c}{$\delta\nu^{-1}$} &
       \multicolumn{1}{c}{$\delta\nu^{+1}$} &&
       \multicolumn{1}{c}{$\delta\nu^{-1}$} &
       \multicolumn{1}{c}{$\delta\nu^{+1}$} &&
       \multicolumn{1}{c}{$\delta\nu^{-1}$} &
       \multicolumn{1}{c}{$\delta\nu^{+1}$} \\
    \hline\\[-0.35cm]
      \multicolumn{3}{l}{c, $v=0$:} \\
        0	&		        &	0.020 234	&&		    &	--0.1	&&		&	1.3	&&		    &	0	    \\
	2	&	0.039 422	&	0.026 643	&&	--0.1	&	--0.1	&&	2.7	&	1.7	&&	--34    &	18  	\\
	4	&	0.033 102	&	0.028 839	&&	--0.1	&	--0.1	&&	2.3	&	1.8	&&	--50    &	37  	\\
	6	&	0.030 957	&	0.030 212	&&	--0.1	&	--0.1	&&	2.2	&	1.9	&&	--68    &	55  	\\
	8	&	0.029 659	&	0.031 288	&&	--0.1	&	--0.1	&&	2.1	&	1.9	&&	--85    &	74  	\\
	10	&	0.028 681	&	0.032 227	&&	--0.1	&	--0.1	&&	2.1	&	2.0	&&	--103	&	93  	\\
    \hline\\[-0.35cm]
      \multicolumn{3}{l}{c, $v=1$:} \\
	0	&		        &	0.020 616	&&		    &	--0.1	&&		&	1.4	&&		    &	0 	    \\
	2	&	0.040 179	&	0.027 149	&&	--0.2	&	--0.1	&&	2.7	&	1.7	&&	--34  	&	19  	\\
	4	&	0.033 742	&	0.029 390	&&	--0.1	&	--0.1	&&	2.3	&	1.9	&&	--51 	&	37  	\\
	6	&	0.031 562	&	0.030 794	&&	--0.1	&	--0.1	&&	2.2	&	1.9	&&	--69 	&	56  	\\
	8	&	0.030 246	&	0.031 899	&&	--0.1	&	--0.1	&&	2.1	&	2.0	&&	--87 	&	76  	\\
	10	&	0.029 258	&	0.032 865	&&	--0.1	&	--0.1	&&	2.1	&	2.0	&&	--105	&	95  	\\
    \hline\\[-0.35cm]
      \multicolumn{3}{l}{c, $v=2$:} \\    
	0	&		        &	0.021 066	&&	    	&	--0.1	&&		&	1.4	&&		    &	0 	    \\
	2	&	0.041 062	&	0.027 755	&&	--0.2	&	--0.1	&&	2.7	&	1.8	&&	--35 	&	19  	\\
	4	&	0.034 481	&	0.030 080	&&	--0.2	& 	 0.7	&&	2.3	&	2.7	&&	--52 	&	39  	\\
	6	&	0.032 254	&	0.031 496	&&	--0.2	&	--0.3	&&	2.2	&	1.8	&&	--71 	&	58  	\\
	8	&	0.030 910	&	0.032 643	&&	--0.2	&	--0.2	&&	2.1	&	1.9	&&	--90 	&	78  	\\
	10	&	0.029 902	&	0.033 653	&&	--0.2	&	--0.2	&&	2.1	&	1.9	&&	--108	&	98  	\\
    \hline\\[-0.35cm]
      \multicolumn{3}{l}{c, $v=3$:} \\
	0	&		        &	0.021 575	&&	    	&	0.0 	&&		&	1.5	&& 	        &	0	    \\
	2	&	0.042 034	&	0.028 451	&&	--0.2	&	0.1 	&&	2.8	&	2.0	&&	--36  	&	21  	\\
	4	&	0.035 328	&	0.030 827	&&	1.3	    &	0.1 	&&	3.8	&	2.2	&&	--53 	&	41  	\\
	6	&	0.033 006	&	0.032 335	&&	0.4	    &	0.1 	&&	2.8	&	2.2	&&	--73 	&	62  	\\
	8	&	0.031 625	&	0.033 538	&&	0.3  	&	0.1 	&&	2.7	&	2.3	&&	--93 	&	83  	\\
	10	&	0.030 575	&	0.034 608	&&	0.3 	&	0.2	    &&	2.6	&	2.3	&&	--113	&	104  	\\
    \hline\\[-0.35cm]
     \multicolumn{3}{l}{c, $v=4$:} \\
	0	&		        &	0.022 045	&&		    &	--0.1	&&		&	1.5	&&		    &	0	    \\
	2	&	0.042 891	&	0.029 123	&&	--0.2	&	--0.1	&&	2.9	&	1.9	&&	--39	&	23  	\\
	4	&	0.035 942	&	0.031 597	&&	--0.2	&	--0.1	&&	2.4	&	2.0	&&	--59	&	45  	\\
	6	&	0.033 546	&	0.033 185	&&	--0.2	&	--0.1	&&	2.3	&	2.1	&&	--80	&	68  	\\
	8	&	0.032 072	&	0.034 461	&&	--0.2	&	0.0	    &&	2.2	&	2.2	&&	--102	&	91  	\\
	10	&	0.030 948	&	0.035 600	&&	--0.2	&	0.0	    &&	2.2	&	2.2	&&	--123	&	115  	\\
    \hline\hline
  \end{tabular}
  }
\end{table}

\clearpage
\section{Simple calculations for the Cartesian and the spherical representations}

The transition from the Cartesian to the spherical representation does not alter the Born-Oppenheimer (BO) and adiabatic energies of the states.
The variational ECG computations were carried out in the $\mathcal{B}(xyz)$ Cartesian form in QUANTEN; however, for solving the rovibronic problem, it is more advantageous to use the spherical representation.

\subsection{Spherical vs. Cartesian relations for the orbital angular momentum couplings and corrections \label{sec:SphCart}}
In this section, we assume the identical $\Sigma$ BF spin projection quantum number in the bra and in the ket unless explicitly stated otherwise.
\begin{align}
  \langle %
    c | \hL_a | c
  \rangle 
  &= 0
  \;, \quad
  \langle %
    b^x | \hL_a | b^x 
  \rangle 
  = 0
  \;, \quad
  \langle %
    b^y | \hL_a | b^y 
  \rangle 
  = 0  \;, \quad a=x,y,z 
  \nonumber \\
  \langle %
    b^x | \hL_x | c
  \rangle 
  &= 0
  \;, \quad
  \langle %
    b^x | \hL_y | c
  \rangle 
  = \iim\eta\ (\eta\in\mathbb{R})
  \;, \quad
  \langle %
    b^x | \hL_z | c
  \rangle 
  = 0
  \nonumber \\
  \langle %
    b^y | \hL_y | c
  \rangle 
  &= 0
  \;, \quad
  \langle %
    b^y | \hL_x | c
  \rangle 
  = -\iim\eta\ (\eta\in\mathbb{R})
  \;, \quad
  \langle %
    b^y | \hL_z | c
  \rangle 
  = 0  
\end{align}

\begin{align}
  \langle 
    b^x | \hL_z | b^y 
  \rangle = -\iim 
  \quad 
  \langle 
    b^y | \hL_z | b^x
  \rangle = +\iim   
\end{align}
\begin{align}
  b^\pm
  =
  \frac{1}{\sqrt{2}}[b^x \pm \iim b^y] \; .
  \label{eq:Pipm}
\end{align}
Then, 
\begin{align}
  \langle 
    b^+ | \hL_z | b^+
  \rangle 
  &=
  \frac{1}{2}
  \langle
    b^x + \iim b^y 
    | \hL_z |
    b^x + \iim b^y 
  \rangle
  \nonumber \\
  &=
  \frac{1}{2}
  \left[%
    \langle
      b^x | \hL_z | b^x 
    \rangle
  -\iim 
    \langle
      b^y | \hL_z | b^x 
    \rangle
  +\iim 
    \langle
      b^x | \hL_z | b^y 
    \rangle
  +
    \langle
      b^y | \hL_z | b^y 
    \rangle  
  \right]
  \nonumber \\
  &=
  \frac{1}{2}
  \left[%
    0
  -\iim 
    \langle
      b^y | \hL_z | b^x 
    \rangle
  +\iim 
    \langle
      b^x | \hL_z | b^y 
    \rangle
  +0
  \right]
  =
  1  
\end{align}

\begin{align}
  \langle 
    b^- | \hL_z | b^-
  \rangle 
  &=
  \frac{1}{2}
  \langle
    b^x - \iim b^y 
    | \hL_z |
    b^x - \iim b^y 
  \rangle
  \nonumber \\
  &=
  \frac{1}{2}
  \left[%
    \langle
      b^x | \hL_z | b^x 
    \rangle
  +\iim 
    \langle
      b^y | \hL_z | b^x 
    \rangle
  -\iim 
    \langle
      b^x | \hL_z | b^y 
    \rangle
  +
    \langle
      b^y | \hL_z | b^y 
    \rangle  
  \right]
  \nonumber \\
  &=
  \frac{1}{2}
  \left[%
    0
   -1
   -1  
   +0
  \right]
  = -1
\end{align}
\begin{align}
  \langle 
    b^+ | \hL_z | b^-
  \rangle 
  &=
  \frac{1}{2}
  \langle
    b^x - \iim b^y 
    | \hL_z |
    b^x + \iim b^y 
  \rangle
  \nonumber \\  
  &=
  \frac{1}{2}
  \left[
    0
    +\iim
    \langle 
      b^y | \hL_z | b^x
    \rangle
    +\iim
    \langle 
      b^x | \hL_z | b^y
    \rangle    
    -0    
  \right]
  =0
\end{align}
\begin{align}
  \langle 
    b^- | \hL_z | b^+
  \rangle 
  &=
  0
\end{align}
\begin{align}
  \langle %
    b^{\pm} | \bos{L} | b^{\pm} %
  \rangle
  =
  \left(%
  \begin{array}{@{}c@{}}
    0 \\ 0 \\ 
    \pm 1
  \end{array}
  \right)
  \quad\text{and}\quad
  \langle %
    b^{\pm} | \bos{L} | b^{\mp} %
  \rangle
  =
  \left(%
  \begin{array}{@{}c@{}}
    0 \\ 0 \\ 0
  \end{array}
  \right) \; .
\end{align}

\begin{align}
  \langle b^x | \hL_x | c \rangle = 0
  \quad\quad
  \langle b^y | \hL_y | c \rangle = 0 
\end{align}
\begin{align}
  \langle b^x | \hL_y | c \rangle = -\iim \eta
  \quad\quad
  \langle b^y | \hL_x | c \rangle = \iim \eta
\end{align}
\begin{align}
  \langle%
    b^+ | \hL^+ c 
  \rangle
  =
  \frac{1}{\sqrt{2}}
  \langle%
    b^x +\iim b^y | [\hL_x + \iim \hL_y] c
  \rangle
  &=
  \frac{1}{\sqrt{2}}
  [%
   \langle b^x | \iim\hL_y | c \rangle
   -\iim
   \langle b^y | \hL_x | c \rangle 
  ]
  \nonumber \\
  &=
  \frac{1}{\sqrt{2}}
  [%
   \iim\langle b^x|\hL_y|c\rangle 
  -\iim\langle b^y|\hL_x|c\rangle
  ]
  \nonumber \\
  &=
  \frac{1}{\sqrt{2}} [\iim (-\iim) \eta - \iim(\iim) \eta]
  =
  \sqrt{2}\eta
\end{align}

\begin{align}
  \langle b^+ | \hat{L}_a^2 | b^+ \rangle
  &=
  \frac{1}{2} \langle b^x+\iim b^y | \hat{L}_a^2 | b^x+\iim b^y \rangle 
  \nonumber \\
  &=
  \frac{1}{2} 
  [%
  \langle b^x | \hat{L}_a^2 | b^x \rangle 
  -\iim \langle b^y | \hat{L}_a^2 | b^x \rangle 
  +\iim \langle b^x | \hat{L}_a^2 | b^y \rangle   
  +(-\iim)\iim \langle b^y | \hat{L}_a^2 | b^y \rangle     
  ]
  \nonumber \\
  &=
  \frac{1}{2} 
  [%
  \langle b^x | \hat{L}_a^2 | b^x \rangle 
  -0
  +0
  + \langle b^y | \hat{L}_a^2 | b^y \rangle     
  ]
  \nonumber \\
  &=
  \frac{1}{2} 
  [%
  \langle b^x | \hat{L}_a^2 | b^x \rangle 
  + \langle b^y | \hat{L}_a^2 | b^y \rangle     
  ]  
\end{align}

\begin{align}
  \langle b^- | \hat{L}_a^2 | b^- \rangle
  &=
  \frac{1}{2} \langle b^x-\iim b^y | \hat{L}_a^2 | b^x-\iim b^y \rangle 
  \nonumber \\
  &=
  \frac{1}{2} 
  [%
  \langle b^x | \hat{L}_a^2 | b^x \rangle 
  +\iim \langle b^y | \hat{L}_a^2 | b^x \rangle 
  -\iim \langle b^x | \hat{L}_a^2 | b^y \rangle   
  +\iim(-\iim) \langle b^y | \hat{L}_a^2 | b^y \rangle     
  ]
  \nonumber \\
  &=
  \frac{1}{2} 
  [%
  \langle b^x | \hat{L}_a^2 | b^x \rangle 
  -0
  +0
  + \langle b^y | \hat{L}_a^2 | b^y \rangle     
  ]
  \nonumber \\
  &=
  \frac{1}{2} 
  [%
  \langle b^x | \hat{L}_a^2 | b^x \rangle 
  + \langle b^y | \hat{L}_a^2 | b^y \rangle     
  ]  
\end{align}

\begin{align}
  \langle b^+ | \hat{L}_a^2 | b^- \rangle
  &=
  \frac{1}{2} \langle b^x+\iim b^y | \hat{L}_a^2 | b^x-\iim b^y \rangle 
  \nonumber \\
  &=
  \frac{1}{2} 
  [%
  \langle b^x | \hat{L}_a^2 | b^x \rangle 
  -\iim \langle b^y | \hat{L}_a^2 | b^x \rangle 
  -\iim \langle b^x | \hat{L}_a^2 | b^y \rangle   
  +(-\iim)^2 \langle b^y | \hat{L}_a^2 | b^y \rangle     
  ]
  \nonumber \\
  &=
  \frac{1}{2} 
  [%
  \langle b^x | \hat{L}_a^2 | b^x \rangle 
  - 
  \langle b^y | \hat{L}_a^2 | b^y \rangle     
  ]
\end{align}

\begin{align}
  \langle b^- | \hat{L}_a^2 | b^+ \rangle
  &=
  \frac{1}{2} \langle b^x-\iim b^y | \hat{L}_a^2 | b^x+\iim b^y \rangle 
  \nonumber \\
  &=
  \frac{1}{2} 
  [%
  \langle b^x | \hat{L}_a^2 | b^x \rangle 
  +\iim \langle b^y | \hat{L}_a^2 | b^x \rangle 
  +\iim \langle b^x | \hat{L}_a^2 | b^y \rangle   
  +(\iim)^2 \langle b^y | \hat{L}_a^2 | b^y \rangle     
  ]
  \nonumber \\
  &=
  \frac{1}{2} 
  [%
  \langle b^x | \hat{L}_a^2 | b^x \rangle 
  - 
  \langle b^y | \hat{L}_a^2 | b^y \rangle     
  ]
\end{align}
Note: $\langle b^x |\hat{L}_x^2 | b^x \rangle = \langle b^y |\hat{L}_y^2 | b^y \rangle$, 
$\langle b^y |\hat{L}_x^2 | b^y \rangle = \langle b^x |\hat{L}_y^2 | b^x \rangle$, so
\begin{align}
  \langle b^+ | \hat{L}_x^2+\hat{L}_y^2 | b^+ \rangle 
  =
  \langle b^- | \hat{L}_x^2+\hat{L}_y^2 | b^- \rangle  
  =
  \langle b^x | \hat{L}^2_x + \hat{L}^2_y | b^x \rangle
  =
  \langle b^y | \hat{L}^2_x + \hat{L}^2_y | b^y \rangle  
\end{align}

\begin{align}
  \langle b^+ | \hat{L}_x^2+\hat{L}_y^2 | b^- \rangle 
  =
  \langle b^- | \hat{L}_x^2+\hat{L}_y^2 | b^+ \rangle 
  =
  0
\end{align}
Similar relations apply for the B (and C) state components.

\clearpage
\subsection{Spherical vs. Cartesian basis relations for the relativistic and QED couplings \label{sec:sphCart2}}

\begin{table}
  \caption{%
    $\mathcal{B}(xyz)$ Cartesian form of the non-vanishing relativistic QED couplings, %
    $\beta_i,\gamma_i,\delta_i,\epsi_i (i=1,2),\lambda,\xi,\zeta\in\mathbb{R}$, within the bcBC electronic-spin subspace of He$_2$ as computed with QUANTEN. 
    The Cartesian component of the spatial part and the $\Sigma$ quantum number of the electron-spin part are in the superscript.
    The `$.$' labels zero (0).
    Note: 
    $\beta_2=-\beta_1/2$ and $\varepsilon_2=-\varepsilon_1/2$; in Ref.~\citenum{paper3}, the $\beta_1=\beta$ and $\epsi_1=\epsi$ labelling is used.
    \label{tab:relcpl}
  }
\scalebox{1.}{%
  \begin{tabular}{@{}|l||ccc @{\ }|@{\ } ccc @{\ }|@{\ } ccc | cc |c |@{}}
    \hline\\[-0.35cm]
    & $\bel^{x,-1}$ & $\bel^{x,0}$ & $\bel^{x,1}$ %
    & $\bel^{y,-1}$ & $\bel^{y,0}$ & $\bel^{y,1}$ 
    & $\cel^{0,-1}$ & $\cel^{0,0}$ & $\cel^{0,1}$
    & $\Bbel^{x,0}$ & $\Bbel^{y,0}$ & $\Ccel^{0,0}$ 
    \\
    \hline\hline\\[-0.35cm]
    $\bel^{x,-1}$
    & $\beta_2$       & .               & $-\gamma_1$       
    & $-\iim\gamma_2$ & .               & $-\iim\gamma_1$   
    & .               & $\delta_2$      & .                 
    & .               & .                                   
    & $\lambda$                                             
    \\ 
    $\bel^{x,0}$
    & .               & $\beta_1$       & .                 
    & .               & .               & .                 
    & $-\delta_1$     & .               & $\delta_1$        
    & .               & $\iim \zeta$                        
    & .                                                     
    \\     
    $\bel^{x,1}$
    & $-\gamma_1$     & .               & $\beta_2$         
    & $\iim\gamma_1$  & .               & $\iim\gamma_2$    
    & .               & $-\delta_2$     & .                 
    & .               & .                                   
    & $\lambda$                                             
    \\
    \hline\\[-0.35cm]
    $\bel^{y,-1}$
    & $\iim\gamma_2$  & .               & $-\iim\gamma_1$   
    & $\beta_2$       & .               & $\gamma_1$        
    & .               & $\iim\delta_2$  & .                 
    & .               & .                                   
    & $\iim \lambda$                                        
    \\ 
    $\bel^{y,0}$
    & .               & .               & .                 
    & .               & $\beta_1$       & .                 
    & $\iim\delta_1$  & .               & $\iim\delta_1$    
    & $-\iim \zeta$   & .                                   
    & .                                                     
    \\     
    $\bel^{y,1}$
    & $\iim\gamma_1$  & .               & $-\iim\gamma_2$   
    & $\gamma_1$      & .               & $\beta_2$         
    & .               & $\iim\delta_2$  & .                 
    & .               & .                                   
    & $-\iim \lambda$                                       
    \\     
    \hline\\[-0.35cm]
    $\cel^{0,-1}$
    & .               & $-\delta_1$     & .                 
    & .               & $-\iim\delta_1$ & .                 
    & $\epsi_2$       & .               & .                 
    & $\xi$           & $\iim\xi$                           
    & .                                                     
    \\ 
    $\cel^{0,0}$
    & $\delta_2$      & .               & $-\delta_2$       
    & $-\iim\delta_2$ & .               & $-\iim\delta_2$   
    & .               & $\epsi_1$       & .                 
    & .               & .                                   
    & .                                                     
    \\ 
    $\cel^{0,1}$ 
    & .               & $\delta_1$      & .                 
    & .               & $-\iim\delta_1$ & .                 
    & .               & .               & $\epsi_2$         
    & $\xi$           & $-\iim\xi$                          
    & .                                                     
    \\ 
    \hline\\[-0.35cm]
    $\Bbel^{x,0}$
    & .               & .               & .                 
    & .               & $\iim\zeta$     & .                 
    & $\xi$           & .               & $\xi$             
    & .               & .                                   
    & .                                                     
    \\         
    $\Bbel^{y,0}$
    & .               & $-\iim\zeta$    & .                 
    & .               & .               & .                 
    & $-\iim\xi$      & .               & $\iim\xi$         
    & .               & .                                   
    & .                                                     
    \\
    \hline\\[-0.35cm]
    $\Ccel^{0,0}$
    & $\lambda $      & .               & $\lambda$         
    & $-\iim\lambda$  & .               & $\iim\lambda$     
    & .               & .               & .                 
    & .               & .                                   
    & .                                                     
    \\         
    \hline 
  \end{tabular}
}
\end{table}

\begin{table}
  \caption{%
    $\mathcal{B}(-1,0,1)$ 
    spherical form of the 
    relativistic QED, $\beta_i,\gamma_i,\epsi_i (i=1,2),\lambda,\xi,\zeta\in\mathbb{R}$ and non-adiabatic, $\eta,\kappa\in\mathbb{R}$ couplings of the $\bel\cel\Bbel\Ccel$ electronic-spin subspace of He$_2$ used in the rovibronic computations. 
    We also show the non-vanishing $\langle\vphipn|\hat{L}^+ \vphi_n\rangle$ couplings ($\eta,\kappa$). 
    All couplings are computed in the Cartesian representation with QUANTEN, the Cartesian to spherical transformation details are in the text.
    Please see also the caption to Table~\ref{tab:relcpl}.
    \label{tab:nadcpl-form}
  }
\scalebox{1.}{%
  \begin{tabular}{@{}|r@{\ }r@{\ }r @{\ \ } l@{\ }||c@{}c@{}c @{\ }|@{\ } c@{}c@{}c @{\ }|@{\ }  c@{}c@{}c@{\ }|@{\ } c@{}c@{\ }|@{\ }c|@{}} 
    \hline\\[-0.40cm]
    \multicolumn{3}{r}{$|\text{ket}\rangle$~~} \\[-0.40cm]
    & & & 
    $\Lambda$
    &  +1 &  +1 &  +1
    & --1 & --1 & --1
    &   0 &   0 &   0
    &  +1 & --1
    &   0
    \\
    & & &
    $\Sigma$
    & --1 & 0 &  +1
    & --1 & 0 &  +1
    & --1 & 0 &  +1
    &   0 & 0
    &   0
    \\   
    & & &
    $\Omega$
    &   0 &  +1 &  +2
    & --2 & --1 &   0
    & --1 &   0 &  +1 
    &  +1 & --1
    &   0
    \\[-0.40cm]
    \multicolumn{11}{l}{$\langle \text{bra}|$} & \\
    $\Lambda$ & $\Sigma$ & $\Omega$ &
    & $\bel^{+,-}$ & $\bel^{+,0}$ & $\bel^{+,+}$ %
    & $\bel^{-,-}$ & $\bel^{-,0}$ & $\bel^{-,+}$ 
    & $\cel^{0,-}$ & $\cel^{0,0}$ & $\cel^{0,+}$
    & $B^{+,0}$    & $B^{-,0}$ 
    & $C^{0,0}$
    \\
    \hline\hline\\[-0.35cm]
      +1 & --1 &   0 & %
    $\bel^{+,-}$
    & $\beta_2$ + $\gamma_2$ & .       & .                               
    & .                      & .       & $-2\gamma_1$                    
    & $\sqrt{2}\eta$ & $\sqrt{2}\delta_2$ & .                            
    & .        & .                                                       
    & $\sqrt{2}\lambda$                                                  
    \\ 
      +1 &   0 &  +1 & %
    $\bel^{+,0}$
    & .        & $\beta_1$       & .                                     
    & .        & .               & .                                     
    & .        & $\sqrt{2}\eta$  & $\sqrt{2}\delta_1$                    
    & $-\zeta$ & .                                                       
    & .                                                                  
    \\ 
      +1 &  +1 &  +2 & %
    $\bel^{+,+}$
    & .        & .       & $\beta_2-\gamma_2$                            
    & .        & .       & .                                             
    & .        & .       & $\sqrt{2}\eta$                                
    & .        & .                                                       
    & .                                                                  
    \\ 
    \hline\\[-0.35cm]
     --1 & --1 & --2 & %
    $\bel^{-,-}$
    & .                   & .       & .                                  
    & $\beta_2-\gamma_2$  & .       & .                                  
    & $-\sqrt{2}\eta$     & .       & .                                  
    & .                   & .                                             
    & .                                                                  
    \\ 
     --1 &   0 & --1 & %
    $\bel^{-,0}$
    & .                   & .                 & .                        
    & .                   & $\beta_1$         & .                        
    & $-\sqrt{2}\delta_1$ & $-\sqrt{2}\eta$   & .                        
    & .        & $\zeta$                                                 
    & .                                                                  
    \\     
     --1 &  +1 &   0 & %
    $\bel^{-,+}$
    & $-2\gamma_1$ & .                   & .                             
    & .            & .                   & $\beta_2 + \gamma_2$          
    & .            & $-\sqrt{2}\delta_2$ & $-\sqrt{2}\eta$               
    & .            & .                                                   
    & $\sqrt{2}\lambda$                                                  
    \\ 
    \hline\\[-0.35cm]
       0 & --1 & --1 & %
    $\cel^{0,-}$
    & $\sqrt{2}\eta$      & .                   & .                      
    & $-\sqrt{2}\eta$     & $-\sqrt{2}\delta_1$ & .                      
    & $\varepsilon_2$     & .                   & .                      
    & .                   & $\sqrt{2}\xi$                                
    & .                                                                  
    \\ 
       0 &   0 &   0 & %
    $\cel^{0,0}$
    & $\sqrt{2}\delta_2$  & $\sqrt{2}\eta$      & .                      
    & .                   & $-\sqrt{2}\eta$     & $-\sqrt{2}\delta_2$    
    & .                   & $\varepsilon_1$     & .                      
    & .                   & .                                            
    & .                                                                  
    \\ 
       0 &  +1 &  +1 & %
    $\cel^{0,+}$ 
    & .        & $\sqrt{2}\delta_1$  & $\sqrt{2}\eta$                    
    & .        & .                   & $-\sqrt{2}\eta$                   
    & .        & .                   & $\varepsilon_2$                   
    & $\sqrt{2}\xi$         & .                                          
    & .                                                                  
    \\ 
    \hline\\[-0.35cm]
    +1  & 0  & +1   & %
    $B^{+,0}$
    & .                   & $-\zeta$            & .                      
    & .                   & .                   & .                      
    &  .                  & .                   & $\sqrt{2}\xi$          
    & .                   & .                                            
    & $\sqrt{2}\kappa$                                                   
    \\     
    --1 & 0  & --1  & %
    $B^{-,0}$
    & .                   & .                   & .                      
    & .                   & $\zeta$             & .                      
    & $\sqrt{2}\xi$       & .                   & .                      
    & .                   & .                                            
    & $-\sqrt{2}\kappa$                                                  
    \\ 
    \hline\\[-0.35cm]
    0 & 0  & 0  & %
    $C^{0,0}$
    & $\sqrt{2}\lambda$   & .                   & .                      
    & .                   & .                   & $\sqrt{2}\lambda$      
    & .                   & .                   & .                      
    & $\sqrt{2}\kappa$    & $-\sqrt{2}\kappa$                            
    & .                                                                  
    \\     
    \hline 
  \end{tabular}
}
\end{table}

In what follows $\hat{H}_{\text{rQ}}$ labels the spin-dependent relativistic and/or leading-order QED operator, $\alpha^2\hat{H}^{(2)}_{\text{sd}}$ or $\alpha^3\hat{H}^{(3)}_{\text{sd}}$ or $\alpha^2\hat{H}^{(2)}_{\text{sd}}+\alpha^3\hat{H}^{(3)}_{\text{sd}}$. Furthermore, it is necessary to write out the BF spin projection quantum number ($\Sigma$) in this section.
The non-vanishing coupling elements in the Cartesian basis representation (as computed with QUANTEN) are collected in Table~\ref{tab:relcpl}.

\begin{align}
  \langle \bx | \hat{H}_{\text{rQ}} | \bx \rangle 
  &=
  \langle \by | \hat{H}_{\text{rQ}} | \by \rangle 
  =: 
  \beta \; , \quad \beta\in\mathbb{R} \\
  \langle \bx | \hat{H}_{\text{rQ}} | \by \rangle 
  &=
  -\langle \by | \hat{H}_{\text{rQ}} | \bx \rangle 
  =:
  \iim \gamma \; , \quad \gamma\in\mathbb{R}
\end{align}

\begin{align}
  &\langle %
    \bpm | \hat{H}_{\text{rQ}} | \bpm 
  \rangle 
  \nonumber \\
  &=
  \frac{1}{2}
  \langle %
    \bx \pm \iim \by | \hat{H}_{\text{rQ}} | \bx \pm \iim \by
  \rangle  
  \nonumber \\
  &=
  \frac{1}{2}
  \left[%
    \langle \bx | \hat{H}_{\text{rQ}} | \bx \rangle
    +
    \langle \by | \hat{H}_{\text{rQ}} | \by \rangle
    \pm \iim
    \langle \bx | \hat{H}_{\text{rQ}} | \by \rangle
    \mp \iim
    \langle \by | \hat{H}_{\text{rQ}} | \bx \rangle
  \right]
  \nonumber \\
  &=
  \frac{1}{2}
  \left[%
    \beta
    +
    \beta
    \pm \iim (\iim\gamma)
    \mp \iim (\iim\gamma)^\ast
  \right]  
  =
  \beta \mp\gamma \in\mathbb{R}
\end{align}

\begin{align}
  &\langle %
    \bmp | \hat{H}_{\text{rQ}} | \bpm 
  \rangle 
  \nonumber \\
  &=
  \frac{1}{2}
  \langle %
    \bx \mp \iim \by | \hat{H}_{\text{rQ}} | \bx \pm \iim \by
  \rangle  
  \nonumber \\
  &=
  \frac{1}{2}
  \left[%
    \langle \bx | \hat{H}_{\text{rQ}} | \bx \rangle
    -
    \langle \by | \hat{H}_{\text{rQ}} | \by \rangle
    \pm \iim
    \langle \bx | \hat{H}_{\text{rQ}} | \by \rangle
    \pm \iim
    \langle \by | \hat{H}_{\text{rQ}} | \bx \rangle
  \right]
  \nonumber \\
  &=
  \frac{1}{2}
  \left[%
    \beta
    -
    \beta
    \pm \iim
    (\iim\gamma)
    \pm \iim
    (\iim\gamma)^\ast
  \right]
  =
  0
\end{align}
We use the conventions introduced in the article, $\hat{O}_a\ (a=x,y,z)$ and $\hat{O}^{\pm}$ are BF operators.
\begin{align}
  B^{+,\Sigma}
  &=
  \frac{1}{\sqrt{2}}\left[B^{x,\Sigma} + \iim B^{y,\Sigma} \right] 
\quad\quad\quad %
  B^{-,\Sigma}
  =
  \frac{1}{\sqrt{2}}\left[B^{x,\Sigma} - \iim B^{y,\Sigma} \right] 
\end{align}
\begin{align}
  \langle b^{x,0} |\hat{H}_{\text{rQ}} | B^{y,0} \rangle = \iim \zeta
  \quad\text{and}\quad
  \langle b^{y,0} |\hat{H}_{\text{rQ}} | B^{x,0} \rangle = -\iim \zeta
\end{align}
In intermediate calculation steps, we drop the $\Sigma$ superscript for brevity,
\begin{align}
  \langle b^{+,0} | \hat{H}_{\text{rQ}} | B^{+,0} \rangle
  &=
  \frac{1}{2}
  \langle b^x + \iim b^y | \hat{H}_{\text{rQ}} | B^x + \iim B^y \rangle
  \nonumber \\
  &=
  \frac{1}{2}[%
  \langle b^x | \hat{H}_{\text{rQ}} | B^x \rangle 
  -\iim 
  \langle b^y | \hat{H}_{\text{rQ}} | B^x \rangle
  +\iim
  \langle b^x | \hat{H}_{\text{rQ}} | B^y \rangle
  +
  \langle b^y | \hat{H}_{\text{rQ}} | B^y \rangle
  ]
  \nonumber \\
  &=
  \frac{1}{2}[%
  0
  -\zeta
  -\zeta
  +
  0
  ] = -\zeta
\end{align}
\begin{align}
  \langle b^{-,0} | \hat{H}_{\text{rQ}} | B^{-,0} \rangle
  &=
  \frac{1}{2}
  \langle b^x - \iim b^y | \hat{H}_{\text{rQ}} | B^x - \iim B^y\rangle
  \nonumber \\
  &=
  \frac{1}{2}
  [%
  \langle b^x | \hat{H}_{\text{rQ}} | B^x \rangle
  +\iim
  \langle b^y | \hat{H}_{\text{rQ}} | B^x \rangle
  -\iim
  \langle b^x | \hat{H}_{\text{rQ}} | B^y \rangle  
  -\iim^2 
  \langle b^y | \hat{H}_{\text{rQ}} | B^y\rangle  
  ]
  \nonumber \\
  &=
  \frac{1}{2}
  [%
  0
  -\iim^2 \zeta
  -\iim^2 \zeta
  +0
  ] = \zeta
\end{align}
\begin{align}
  \langle b^{+,0} | \hat{H}_{\text{rQ}} | B^{-,0} \rangle
  &=
  \frac{1}{2}
  \langle b^x + \iim b^y | \hat{H}_{\text{rQ}} | B^x - \iim B^y \rangle
  \nonumber \\
  &=
  \frac{1}{2}
  [%
  \langle b^x | \hat{H}_{\text{rQ}} | B^x \rangle
  -\iim 
  \langle b^y | \hat{H}_{\text{rQ}} | B^x \rangle
  -\iim
  \langle b^x | \hat{H}_{\text{rQ}} | B^y \rangle
  -\iim(-\iim)
  \langle b^y | \hat{H}_{\text{rQ}} | B^y \rangle
  ]
  \nonumber \\  
  &=
  \frac{1}{2}
  [%
  0
  -\zeta
  +\zeta
  -
  0
  ] = 0
\end{align}
\begin{align}
  \langle b^{-,0} | \hat{H}_{\text{rQ}} | B^{+,0} \rangle
  &=
  \frac{1}{2}
  \langle b^x - \iim b^y | \hat{H}_{\text{rQ}} | B^x + \iim B^y \rangle 
  \nonumber \\
  &=
  \frac{1}{2}
  [%
  \langle b^x | \hat{H}_{\text{rQ}} | B^x \rangle
  + \iim 
  \langle b^y | \hat{H}_{\text{rQ}} | B^x \rangle
  + \iim
  \langle b^x | \hat{H}_{\text{rQ}} | B^y \rangle    
  + \iim^2
  \langle b^y | \hat{H}_{\text{rQ}} | B^y \rangle
  ]
  \nonumber \\
  &=
  \frac{1}{2}
  [%
  0
  - \iim^2 \zeta
  + \iim^2 \zeta
  + 
  0
  ] = 0 
\end{align}

\begin{align}
  \langle b^{x,-}|\hat{H}_{\text{rQ}}|C^{0,0} \rangle
  =
  \langle b^{x,+}|\hat{H}_{\text{rQ}}|C^{0,0} \rangle
  =
  \lambda
\end{align}
\begin{align}
  \langle b^{y,-}|\hat{H}_{\text{rQ}}|C^{0,0} \rangle = \iim \lambda
  \quad \text{and} \quad
  \langle b^{y,+}|\hat{H}_{\text{rQ}}|C^{0,0} \rangle = -\iim \lambda  
\end{align}
\begin{align}
  \langle b^{+,-} | \hat{H}_{\text{rQ}} | C^{0,0} \rangle
  =
  \frac{1}{\sqrt{2}}
  [%
  \langle b^{x,-} | \hat{H}_{\text{rQ}} | C^{0,0} \rangle
  -\iim
  \langle b^{y,-} | \hat{H}_{\text{rQ}} | C^{0,0} \rangle
  ]
  =
  \frac{1}{\sqrt{2}}
  [%
  \lambda
  -\iim^2 \lambda
  ] = \sqrt{2}\lambda
\end{align}
\begin{align}
  \langle b^{+,+} | \hat{H}_{\text{rQ}} | C^{0,0} \rangle
  =
  \frac{1}{\sqrt{2}}
  [%
  \langle b^{x,+} | \hat{H}_{\text{rQ}} | C^{0,0} \rangle
  -\iim
  \langle b^{y,+} | \hat{H}_{\text{rQ}} | C^{0,0} \rangle
  ]
  =
  \frac{1}{\sqrt{2}}
  [%
  \lambda
  +\iim^2 \lambda
  ] = 0
\end{align}
\begin{align}
  \langle b^{-,-} | \hat{H}_{\text{rQ}} | C^{0,0} \rangle
  =
  \frac{1}{\sqrt{2}}
  [%
  \langle b^{x,-} | \hat{H}_{\text{rQ}} | C^{0,0} \rangle
  +\iim
  \langle b^{y,-} | \hat{H}_{\text{rQ}} | C^{0,0} \rangle
  ]
  =
  \frac{1}{\sqrt{2}}
  [%
  \lambda
  +\iim^2 \lambda
  ] = 0
\end{align}
\begin{align}
  \langle b^{-,+} | \hat{H}_{\text{rQ}} | C^{0,0} \rangle
  =
  \frac{1}{\sqrt{2}}
  [%
  \langle b^{x,+} | \hat{H}_{\text{rQ}} | C^{0,0} \rangle
  +\iim
  \langle b^{y,+} | \hat{H}_{\text{rQ}} | C^{0,0} \rangle
  ]
  =
  \frac{1}{\sqrt{2}}
  [%
  \lambda
  -\iim^2 \lambda
  ] 
  = \sqrt{2} \lambda
\end{align}

\begin{align}
  \langle c^{0,-} |\hat{H}_{\text{rQ}}| B^{x,0} \rangle = 
  \langle c^{0,+} |\hat{H}_{\text{rQ}}| B^{x,0} \rangle = \xi 
  \quad\quad
  \langle c^{0,-} |\hat{H}_{\text{rQ}}| B^{y,0} \rangle = \iim\xi
  \quad\quad
  \langle c^{0,+} |\hat{H}_{\text{rQ}}| B^{y,0} \rangle = -\iim\xi
\end{align}
\begin{align}
  \langle c^{0,-} |\hat{H}_{\text{rQ}}| B^{+,0} \rangle 
  &= 
  \frac{1}{\sqrt{2}}
  \langle c^{0,-} |\hat{H}_{\text{rQ}}| B^{x,0} + \iim B^{y,0} \rangle 
  \nonumber \\
  &=
  \frac{1}{\sqrt{2}}
  [%
  \langle c^{0,-} |\hat{H}_{\text{rQ}}| B^{x,0} \rangle 
  +\iim %
  \langle c^{0,-} |\hat{H}_{\text{rQ}}| B^{y,0} \rangle 
  ]
  \nonumber \\
  &=
  \frac{1}{\sqrt{2}}
  [%
  \xi
  +\iim %
  (\iim)\xi
  ] = 0
\end{align}
\begin{align}
  \langle c^{0,-} |\hat{H}_{\text{rQ}}| B^{-,0} \rangle 
  &= 
  \frac{1}{\sqrt{2}}
  \langle c^{0,-} |\hat{H}_{\text{rQ}}| B^{x,0} - \iim B^{y,0} \rangle 
  \nonumber \\
  &= 
  \frac{1}{\sqrt{2}}
  [%
  \langle c^{0,-} |\hat{H}_{\text{rQ}}| B^{x,0} \rangle 
  -\iim 
  \langle c^{0,-} |\hat{H}_{\text{rQ}}| B^{y,0} \rangle   
  ]
  \nonumber \\
  &= 
  \frac{1}{\sqrt{2}}
  [%
  \xi
  -\iim 
  (\iim)\xi
  ] = \sqrt{2} \xi
\end{align}
\begin{align}
  \langle c^{0,+} |\hat{H}_{\text{rQ}}| B^{+,0} \rangle 
  &= 
  \frac{1}{\sqrt{2}}
  \langle c^{0,+} |\hat{H}_{\text{rQ}}| B^{x,0} + \iim B^{y,0} \rangle 
  \nonumber \\
  &= 
  \frac{1}{\sqrt{2}}
  [%
  \langle c^{0,+} |\hat{H}_{\text{rQ}}| B^{x,0} \rangle 
  +\iim 
  \langle c^{0,+} |\hat{H}_{\text{rQ}}| B^{y,0} \rangle   
  ]
  \nonumber \\
  &= 
  \frac{1}{\sqrt{2}}
  [%
  \xi
  +\iim(-\iim) \xi
  ] = \sqrt{2}\xi
\end{align}
\begin{align}
  \langle c^{0,+} |\hat{H}_{\text{rQ}}| B^{-,0} \rangle 
  &= 
  \frac{1}{\sqrt{2}}
  \langle c^{0,+} |\hat{H}_{\text{rQ}}| B^{x,0} - \iim B^{y,0} \rangle 
  \nonumber \\
  &= 
  \frac{1}{\sqrt{2}}
  [%
  \langle c^{0,+} |\hat{H}_{\text{rQ}}| B^{x,0} \rangle 
  -\iim 
  \langle c^{0,+} |\hat{H}_{\text{rQ}}| B^{y,0} \rangle   
  ]
  \nonumber \\
  &= 
  \frac{1}{\sqrt{2}}
  [%
  \xi
  -\iim (-\iim) \xi
  ] = 0
\end{align}

\clearpage
\section{Auxiliary calculations for the coupled rovibronic equation}
\begin{align}
    \langle
    \vphi_{n'} \DMJp
    |
    \bos{\Delta}_{\brho}
    |
    \vphi_n  \DMJ
    \rangle
    \frac{1}{\rho} g_k
    &=
    \delta_{\ome'\ome}
    \langle
    \vphipn
    | \bos{\Delta}_{\brho}
    | \vphi_n
    \rangle
    \frac{1}{\rho} g_k
    +
    \delta_{n'n}
    \langle 
    \DMJp
    | \bos{\Delta}_{\brho}
    | \DMJ
    \rangle
    \frac{1}{\rho} g_k
    \nonumber \\
    &+
    \delta_{\ome'\ome}
    \delta_{n'n}
    \bos{\Delta}_{\brho}
    \left( 
    \frac{1}{\rho} g_k
    \right)
    +
    2
    \langle \vphi_{n'}
    | \bos{\nabla}_{\brho}
    | \vphi_n
    \rangle
    \langle \DMJp
    | \bos{\nabla}_{\brho}
    | \DMJ
    \rangle
    \frac{1}{\rho} g_k
    \nonumber \\
    &+
    2
    \delta_{\ome'\ome}
    \langle \vphi_{n'}
    | \bos{\nabla}_{\brho}
    | \vphi_n
    \rangle
    \bos{\nabla}_{\brho}
    \left( 
    \frac{1}{\rho} g_k
    \right)
    \nonumber \\
    &+
    2
    \delta_{n'n}
    \langle \DMJp
    | \bos{\nabla}_{\brho}
    | \DMJ
    \rangle
    \bos{\nabla}_{\brho}
    \left( 
    \frac{1}{\rho} g_k
    \right)
    \label{eq:som-nuc-Lap-1}
\end{align}
To evaluate the first three terms, we will use 
\begin{align}
    \bos{\Delta}_{\brho}
    =&
    \frac{1}{\rho^2}
    \frac{\partial}{\partial \rho}
    \left(
    \rho^2
    \frac{\partial}{\partial \rho}
    \right)
    -
    \frac{1}{\rho^2}
    \hat{R}^2
    =
    \frac{1}{\rho^2}
    \frac{\partial}{\partial \rho}
    \left(
    \rho^2
    \frac{\partial}{\partial \rho}
    \right)
    -
    \frac{1}{\rho^2}
    (
    \hbos{J}
    -
    \hbos{L}
    -\hbos{S}
    )^2
    \nonumber \\
    =&
    \frac{1}{\rho^2}
    \frac{\partial}{\partial \rho}
    \left(
    \rho^2
    \frac{\partial}{\partial \rho}
    \right)
    -
    \frac{1}{\rho^2}
    \Big[
    (\hat{J}^2 - J_z^2)
    +
    (\hat{L}^2_x + \hat{L}^2_y)
    +
    (\hat{S}^2 - S_z^2)
    \nonumber \\
    &-
    (\hat{J}^+\hat{S}^- + \hat{J}^-\hat{S}^+)
    -
    (\hat{J}^+\hat{L}^- + \hat{J}^-\hat{L}^+)
    +
    (\hat{S}^+\hat{L}^- + \hat{S}^-\hat{L}^+)
    \Big]
    \; ,
\end{align}
where we exploited that for the diatom's $\bhat{R}$ rotational angular momentum $z$ component vanishes in the BF frame, 
\begin{align}
    0 %
    &= \hat{R}_z^2 %
    = (\hat{J}_z-\hat{L}_z-\hat{S}_z)^2 %
    = \hat{J}_z^2 + \hat{L}_z^2 + \hat{S}_z^2 
    -2 \hat{J}_z \hat{L}_z 
    -2 \hat{J}_z \hat{S}_z 
    +2 \hat{L}_z \hat{S}_z 
    \nonumber \\
    &\Rightarrow\ 
    \hat{J}_z^2 + \hat{L}_z^2 + \hat{L}_z^2 
    =
     2 \hat{J}_z \hat{L}_z 
    +2 \hat{J}_z \hat{S}_z 
    -2 \hat{L}_z \hat{S}_z  \; ,
\end{align}
and thus, we can write,
\begin{align}
    (
    \hbos{J}
    -
    \hbos{L}
    -\hbos{S}
    )^2
    =&
    \hat{J}^2 + \hat{L}^2 + \hat{S}^2 
  -2\hat{\bos{J}} \cdot \hat{\bos{L}} 
  -2\hat{\bos{J}} \cdot \hat{\bos{S}} 
  +2\hat{\bos{L}} \cdot \hat{\bos{S}} 
  \nonumber \\
  =&
  \hat{J}^2 + \hat{L}_x^2 + \hat{L}_y^2 + \hat{L}_z^2 + \hat{S}^2
  -
  2 \hat{J}_z \hat{L}_z 
    -2 \hat{J}_z \hat{S}_z 
    +2 \hat{L}_z \hat{S}_z
  \nonumber \\
  &-
  2 \hat{J}_x \hat{L}_x 
    -2 \hat{J}_x \hat{S}_x 
    +2 \hat{L}_x \hat{S}_x
  -
  2 \hat{J}_y \hat{L}_y 
    -2 \hat{J}_y \hat{S}_y 
    +2 \hat{L}_y \hat{S}_y
    \nonumber \\
    =&
    (\hat{J}^2 - \hat{J}_z^2)
    +
    (\hat{L}^2_x + \hat{L}^2_y)
    +
    (\hat{S}^2 - \hat{S}_z^2)
    \nonumber \\
    &-
    (\hat{J}^+\hat{S}^- + \hat{J}^-\hat{S}^+)
    -
    (\hat{J}^+\hat{L}^- + \hat{J}^-\hat{L}^+)
    +
    (\hat{S}^+\hat{L}^- + \hat{S}^-\hat{L}^+)
    \; .
\end{align}
For the last three terms of Eq.~\eqref{eq:som-nuc-Lap-1}, we can rely on the following relations,
\begin{align}
    \langle a'
    |\bos{\nabla}_{\brho} a\rangle
    \cdot
    \langle b'
    |\bos{\nabla}_{\brho} b \rangle
    &=
    \langle a'
    |\partial_\rho a \rangle
    \cdot
    \langle b'
    |\partial_\rho b \rangle
    -
    \frac{1}{\rho^2}
    \left[
    \langle a'
    |\hat{R}_x a \rangle
    \cdot
    \langle b'
    |\hat{R}_x b \rangle
    +
    \langle a'
    |\hat{R}_y a \rangle
    \cdot
    \langle b'
    |\hat{R}_y b \rangle
    \right]
    \nonumber \\
    &=
    \langle a'
    |\partial_\rho a \rangle
    \cdot
    \langle b'
    |\partial_\rho b \rangle
    -
    \frac{1}{\rho^2}
    \Big[
    \langle a'
    |(\hat{J}_x-\hat{L}_x-\hat{S}_x) a \rangle
    \cdot
    \langle b'
    |(\hat{J}_x-\hat{L}_x-\hat{S}_x) b \rangle
    \nonumber \\
    &+
    \langle a'
    |(\hat{J}_y-\hat{L}_y-\hat{S}_y) a \rangle
    \cdot
    \langle b'
    |(\hat{J}_y-\hat{L}_y-\hat{S}_y) b \rangle
    \Big]
    \nonumber \\
    &=
    \langle a'
    |\partial_\rho a \rangle
    \cdot
    \langle b'
    |\partial_\rho b \rangle
    -
    \frac{1}{2\rho^2}
    \Big[
    \langle a'
    |(\hat{J}^+-\hat{L}^+-\hat{S}^+) a \rangle
    \cdot
    \langle b'
    |(\hat{J}^--\hat{L}^--\hat{S}^-) b \rangle
    \nonumber \\
    &+
    \langle a'
    |(\hat{J}^--\hat{L}^--\hat{S}^-) a \rangle
    \cdot
    \langle b'
    |(\hat{J}^+-\hat{L}^+-\hat{S}^+) b \rangle
    \Big]
    \; ,
\end{align}
where $a$ and $b$ can be any pairing of electronic, rotational, or vibrational functions, where $\hat{R}_z=0$ was used.

Finally, we reiterate the action and matrix elements of the various angular momentum operators,
\begin{align}
    \hat{L}_z \vphi_n
    &=
    \lam \vphi_n
    \; \quad
    \hat{S}_z \vphi_n
    =
    \sig \vphi_n
    \; ,
    \quad\text{and}\quad
    \hat{S}^2 \vphi_n
    =
    S(S+1) \vphi_n
    \; ,
    \nonumber \\
    \hat{J}_z \DMJ
    &=
    \ome \DMJ
    \; ,
    \quad
    \hat{J}_Z \DMJ
    =
    M_J \DMJ
    \; ,
    \quad\text{and}\quad
    \hat{J}^2 \DMJ
    =
    J(J+1) \DMJ
    \; ,
    \nonumber \\
    \langle 
      \DMJp |
      \hat{J}^\pm 
      \DMJ
    \rangle
    &=
    \delta_{\Omega',\Omega \mp 1}
    [J(J+1) - \Omega(\Omega \mp 1)]^{\frac{1}{2}}
    =
    \delta_{\Omega',\Omega \mp 1}
    C_{J\ome}^\mp
    \; , \\
    \langle 
      \vphipn | 
        \hat{S}^\pm
      \vphi_n
    \rangle
    &=
    \delta_{\Sigma',\Sigma \pm 1}
    [S(S+1) - \Sigma (\Sigma \pm 1)]^{\frac{1}{2}}
    =
    \delta_{\Sigma',\Sigma \pm 1}
    C_{S\sig}^\pm
    \; ,
\end{align}
with $\langle \vphi_{n'} | \hat{L}^{\pm} \vphi_n \rangle$ computed numerically, and we note that $\frac{\partial}{\partial \rho} | \DMJ \rangle = 0 $. So, we can proceed with the evaluation of each term in Eq.~\eqref{eq:som-nuc-Lap-1} as
\begin{align}
    \langle
    \vphipn
    | \bos{\Delta}_{\brho}
    | \vphi_n
    \rangle
    &=
    \langle
    \vphipn
    | \frac{1}{\rho^2}
    \frac{\partial}{\partial \rho}
    \left(
    \rho^2
    \frac{\partial}{\partial \rho}
    \right)
    |
     \vphi_n
    \rangle
    -
    \frac{1}{\rho^2}
    \langle
    \vphipn
    |
    (\hat{L}^2_x + \hat{L}^2_y)
    +
    (\hat{S}^2 - S_z^2)
    |
    \vphi_n
    \rangle
    \nonumber \\
    &-
    \frac{1}{\rho^2}
    \langle
    \vphipn
    |
    (\hat{S}^+\hat{L}^- + \hat{S}^-\hat{L}^+)
    \vphi_n
    \rangle
    \nonumber \\
    &=
    -
    \langle
    \frac{\partial}{\partial \rho}
    \vphipn
    |\frac{\partial}{\partial \rho}
    \vphi_n
    \rangle 
    -
    \frac{1}{\rho^2}
    \langle
    \vphipn
    |
    \hat{L}^2_x + \hat{L}^2_y
    |
    \vphi_n
    \rangle
    -
    \delta_{n'n}
    \frac{1}{\rho^2}
    [S(S+1)- \sig^2]
    \nonumber \\
    &-
    \frac{1}{\rho^2}
    \left[ 
    \delta_{\lam'\lam+1}
    \delta_{\sig'\sig+1}
    C_{S\sig}^+
    \langle \vphi_{n'} |
    \hat{L}^+ \vphi_n
    \rangle
    +
    \delta_{\lam'\lam-1}
    \delta_{\sig'\sig-1}
    C_{S\sig}^-
    \langle \vphi_{n'} |
    \hat{L}^- \vphi_n
    \rangle
    \right]
\end{align}
\begin{align}
    \langle \DMJp
    |
    \bos{\Delta}_{\brho}
    | \DMJ
    \rangle
    &=
    -
    \frac{1}{\rho^2}
    \langle \DMJp
    |
   \hat{J}^2 - \hat{J}_z^2
    | \DMJ
    \rangle
    =
    -
    \frac{1}{\rho^2}
    \delta_{\ome'\ome}
    [J(J+1)-\ome^2]
\end{align}
\begin{align}
    \bos{\Delta}_{\brho}
    \left( 
    \frac{1}{\rho} g_k
    \right)
    =
    \frac{1}{\rho^2}
    \frac{\partial}{\partial \rho}
    \left(
    \rho^2
    \frac{\partial}{\partial \rho}
    \right)
    \left[
    \frac{1}{\rho}
    g_k
    \right]
    =
    \frac{1}{\rho}
    \frac{\partial^2 g_k}{\partial \rho^2}
\end{align}
\begin{align}
    2
    \langle \vphi_{n'}
    | \bos{\nabla}_{\brho}
     \vphi_n
    \rangle
    \langle \DMJp
    | \bos{\nabla}_{\brho}
     \DMJ
    \rangle
    &=
    \frac{1}{\rho^2}
    \langle \vphi_{n'}
    | (\hat{S}^+ + \hat{L}^+ )
     \vphi_n
    \rangle
    \langle \DMJp
    | \hat{J}^-
     \DMJ
    \rangle
    \nonumber \\
    &+
    \frac{1}{\rho^2}
    \langle \vphi_{n'}
    | (\hat{S}^- + \hat{L}^- )
     \vphi_n
    \rangle
    \langle \DMJp
    | \hat{J}^+
     \DMJ
    \rangle
    \nonumber \\
    &=
    \frac{1}{\rho^2}
    \Big[
    \delta_{\lam'\lam}
    \delta_{\sig'\sig+1}
    C_{J\ome}^+
    C_{S\sig}^+
    +
    \delta_{\lam'\lam+1}
    \delta_{\sig'\sig}
    C_{J\ome}^+
    \langle \vphi_n |
    \hat{L}^+ \vphipn \rangle
    \nonumber \\
    &+
    \delta_{\lam'\lam}
    \delta_{\sig'\sig-1}
    C_{J\ome}^-
    C_{S\sig}^-
    +
    \delta_{\lam'\lam-1}
    \delta_{\sig'\sig}
    C_{J\ome}^-
    \langle \vphi_n |
    \hat{L}^- \vphipn \rangle
    \Big]
\end{align}
\begin{align}
    2
    \langle \vphi_{n'}
    | \bos{\nabla}_{\brho} %
     \vphi_n
    \rangle
    \bos{\nabla}_{\brho}
    \left( 
    \frac{1}{\rho} g_k
    \right)
    &=
    2
    \langle \vphi_{n'}
    | 
    \frac{\partial}{\partial \rho} %
    \vphi_n
    \rangle
    \cdot
    \frac{\partial}{\partial \rho}
    \left( 
    \frac{1}{\rho} g_k
    \right)
    \label{eq:som-el-vib-cpl}
\end{align}
\begin{align}
    2
    \langle \DMJp
    | \bos{\nabla}_{\brho} %
    \DMJ
    \rangle
    \bos{\nabla}_{\brho}
    \left( 
    \frac{1}{\rho} g_k
    \right)
    = 0
\end{align}
By noting that Eq.~\eqref{eq:som-el-vib-cpl} vanishes for the \btPg\ and \ctSgp\ states, we arrive at the coupled radial equation.
\begin{align}
    \bos{\Delta}_{\brho}
    &=
    \frac{1}{\rho^2}
    \frac{\partial}{\partial \rho}
    \left(
    \rho^2
    \frac{\partial}{\partial \rho}
    \right)
    -
    \frac{1}{\rho^2}
    \hat{\bos{R}}^2
    \nonumber \\    
    &=
    \frac{1}{\rho^2}
    \frac{\partial}{\partial \rho}
    \left(
    \rho^2
    \frac{\partial}{\partial \rho}
    \right)
    \nonumber \\
    &
    -
    \frac{1}{\rho^2}
    \Big[
    (\hat{N}^2 - N_z^2)
    +
    (\hat{L}^2_x + \hat{L}^2_y)
    -
    (N^+L^- + N^-L^+)
    \Big]
    \; .
\end{align}